\documentclass[aps,pre,floatfix,twocolumn,10pt,superscriptaddress,showpacs]%
{revtex4-1}
\pdfoutput=1
\usepackage{graphicx}
\usepackage{amssymb}
\usepackage{natbib}
\usepackage{dcolumn}
\usepackage{bm}
\usepackage{color}
\usepackage{subfigure}
\usepackage{graphicx,psfrag,xspace}
\usepackage{color}
\usepackage{amssymb,amsfonts,amsmath}
\usepackage{setspace}
\usepackage{etoolbox}
\usepackage[T1]{fontenc}
\usepackage{lmodern}

\usepackage{hyperref}
\hypersetup{%
colorlinks=true,%
citecolor=blue,%
filecolor=blue,%
linkcolor=blue,%
urlcolor=blue,%
pdfauthor={V. H. Purrello, J. L. Iguain, A. B. Kolton, E. A. Jagla},%
pdftitle={Creep and thermal rounding close to the elastic depinning threshold}%
}

\begin{document}
\author{V. H. Purrello}
\email{vpurrello@ifimar-conicet.gob.ar}
\author{J. L. Iguain}
\email{iguain@mdp.edu.ar}
\affiliation{Instituto de Investigaciones F\'isicas de Mar del Plata (IFIMAR),
CONICET and Facultad de Ciencias Exactas y Naturales,
Universidad Nacional de Mar del Plata,
De\'an Funes 3350, (7600) Mar del Plata, Argentina}

\author{A. B. Kolton}
\email{koltona@cab.cnea.gov.ar}
\author{E. A. Jagla}
\email{jagla@cab.cnea.gov.ar}
\affiliation{Comisi\'on Nacional de Energ\'ia At\'omica,
Instituto Balseiro (UNCu), and CONICET\\
Centro At\'omico Bariloche, (8400) Bariloche, Argentina}

\title{Creep and thermal rounding close to the elastic depinning threshold}

\begin{abstract}
We study the slow stochastic dynamics near the depinning threshold of an elastic
interface in a random medium by solving a particularly suited model of hopping
interacting particles that belongs to the quenched-Edwards-Wilkinson depinning
universality class. The model allows us to compare the cases of uniformly
activated and Arrhenius activated hops. In the former case, the velocity
accurately follows a standard scaling law of the force and noise intensity with
the analog of the thermal rounding exponent satisfying a modified
``hyperscaling'' relation. For the Arrhenius activation, we find, both
numerically and analytically, that the standard scaling form fails for any value
of the thermal rounding exponent. We propose an alternative scaling
incorporating logarithmic corrections that appropriately fits the numerical
results. We argue that this anomalous scaling is related to the strong
correlation between activated hops that, alternated with deterministic
depinning-like avalanches, occur below the depinning threshold. We rationalize
the spatiotemporal patterns by making an analogy of the present model in the
near-threshold creep regime with some well-known models with extremal dynamics,
particularly the Bak-Sneppen model.
\end{abstract}


\maketitle

\tableofcontents

\section{Introduction}

The understanding of the behavior of elastic lines or surfaces evolving on
disordered potentials under the driving of external forces is of great practical
importance in a variety of fields, as for instance the movement of domain walls
in ferromagnetic materials~\cite{Ferre2013651,ferro1,ferro2,Durin_prl2016},
wetting fronts on a rough substrate~\cite{wet1,ledoussal_epl2009}, seismic
fault dynamics~\cite{eq1,eq2,eq3}, and even in the advance of
reaction~\cite{Atis_prl2015} and cell migration
fronts~\cite{Chepizhko_pnas2016}. The main feature of the dynamics of this kind
of process is the existence of a depinning transition as a function of the
applied driving force $f$. For $f$ lower than a certain critical value $f_c$,
the elastic surface is pinned, and its velocity $v$ is zero. When $f>f_c$, the
interface enters a moving regime with a finite average velocity. The transition
at $f_c$ has many features that allow a description similar to that of
equilibrium critical phenomena~\cite{fisher83,fisher85,fisher,kardar}. In
particular, when approaching $f_c$ from above, the velocity of the interface
plays the role of the order parameter and vanishes, at zero temperature, as a
power law, namely
\begin{equation}
v(f)\sim (f-f_c)^\beta,
\label{eq:v_f_beta}
\end{equation}
where $\beta$ defines the depinning exponent.

The transition from the regime of pinned metastable states to the regime of
moving steady state at $f_c$ is sharp only in the ideal case in which activation
effects are absent. A temporally fluctuating random external field induces
motion and thus a finite velocity of the interface even for $f<f_c$. When $f$ is
much smaller than $f_c$, the induced velocity is usually extremely small because
the effective barriers between metastable states typically diverge as $f \to 0$.
This scenario is known as the creep regime~\cite{ioffe,feigel,natter,chauve}
and is related to the general glassy nature of the ground state of elastic
manifolds in random media. When $f$ is close to $f_c$, the external fluctuating
field produces a smearing of the depinning
transition~\cite{middleton,bustingorry1,bustingorry2,chen,nowak,roters,
vandembroucq}, analogous to the smearing of the magnetization by an external
applied field at a continuous ferromagnetic transition. In particular, right at
$f=f_c$, the velocity of the interface is expected, by analogy with standard
phase transitions, to be a power of the external field amplitude, which we will
note $T$ (although it is not necessarily a temperature), namely
\begin{equation}
v(f_c,T)\sim T^{\psi}
\label{eq:v_t_psi}
\end{equation}
where $\psi$ defines the rounding exponent. In the case the noise is associated
to thermal fluctuations, we speak of $\psi$ as the \textit{thermal rounding
exponent}. Allowing for an applied force that is close, but not exactly equal to
the critical force at a finite (but small) field amplitude $T$, the general
expected scaled form for the velocity as a function of the control parameters
$T$ and $f$ can be written as
\begin{equation}
v(\Delta,T)\sim T^\psi g\left (\Delta/T^{\psi/\beta}\right),
\label{eq:v_t_f}
\end{equation}
where we have defined $\Delta\equiv f-f_c$. The scaling function $g$ has the
limiting behavior $g(0)=1$, and $g(x)\sim x^\beta$ for large $x$, in such a way
that Eqs.~(\ref{eq:v_f_beta}) and (\ref{eq:v_t_psi}) are limiting cases of
Eq.~(\ref{eq:v_t_f}).

The exponent $\psi$ is the main parameter characterizing the rounding effect of
the external noise on the depinning transition, and there have been a number of
attempts to asses its universality and determine its precise value.
Theoretically, a few scaling expressions have been
proposed~\cite{narayan,roters} that relate $\psi$ to other well-known exponents
of the depinning transition. It seems that no solid support exists for these
expressions, however. Moreover, it has been also argued that $\psi$ might be
nonuniversal or less universal than the other depinning exponents, due to its
connection with localized soft modes right at the transition, which are sensible
to microscopic characteristics of the pinning potential~\cite{middleton}. On the
other hand, although mean field results~\cite{fisher83,fisher85} and functional
renormalization group techniques have been successfully applied to the creep
regime for $f\ll f_c$ and to the zero temperature depinning transition, the
description of the thermal rounding regime remains elusive~\cite{chauve2000}.
Numerically, there is also a variety of
results~\cite{chen,nowak,roters,vandembroucq,bustingorry1,bustingorry2}. In
particular, the value of $\psi\simeq 0.15$ found for an elastic string with
short-range elasticity in a random-bond short-correlated pinning
potential~\cite{bustingorry1,bustingorry2} seems to be compatible with
experimental studies of thermal rounding at the depinning transition of domain
walls in thin ferromagnetic films~\cite{metaxas1,metaxas2}. Both the numerical
and experimental determinations of $\psi$ are subtle, however, and the agreement
must be taken with caution. On one hand, power-law corrections to scaling were
shown to be important in one dimension, being the source of large numerical
discrepancies in the literature for the $T=0$ depinning
exponents~\cite{exponentes}. The magnitude of the possible bias these
corrections may induce in the $\psi$ exponent are currently unknown.
Experimentally, on the other hand, although thin ferromagnetic films have become
paradigmatic systems to study universal creep phenomena for $f \ll f_c$,
understanding the effect of temperature right at or very close to $f_c$ remains
a challenge, mainly because it is hard to get a precise estimate for $f_c$ at
finite $T$~\cite{privatejeudy}. The precision achieved so far for $\psi$
numerically or experimentally is thus not enough yet to test the various
theoretical predictions, nor to accurately test the scaling form of
Eq.~(\ref{eq:v_t_f}). A very recent experimental work shows however good
agreement with the scaling form of Eq.~(\ref{eq:v_t_f}) above the depinning
threshold $f>f_c$~\cite{jeudy_PRL_enviado}. This context motivates further
research and new ways to approach the problem of the thermal rounding at the
depinning transition and the dynamics just below the threshold.

Here we study a particular version of the one-di\-men\-sional depinning problem
in the case the pinning potential is composed of very narrow and uncorrelated
pinning wells. The characteristics of the potential make the model particularly
suitable to numerical simulation and to include a  precise scaling analysis. In
particular, it allows us to consider two types of activations from the pinning
wells below the depinning threshold: the usual Arrhenius activation rate,
physically associated to thermal fluctuations, and a uniform activation rate,
which is independent of the height of the energy barriers. We find that in the
case of a uniform activation our results follow closely Eq.~(\ref{eq:v_t_f}),
with a numerical value of $\psi$ that can be linked with other
depinning exponents of the model, unveiling an hyperscaling-like relation. On
the other hand, in the case of a thermal activation, we find results that
\textit{cannot} be appropriately scaled according to Eq.~(\ref{eq:v_t_f}), for
any value of $\psi$. We argue that the data can be rationalized instead
including logarithmic corrections, allowing Eq.~(\ref{eq:v_t_f}) to describe
both the motion just below and above the depinning threshold. Interestingly,
just below the depinning threshold we find that activated events are spatially
and temporally correlated, forming large clusters analogous to depinning
avalanches. This is remarkably similar to what was recently found in creep
simulations well below the threshold~\cite{ferrero_correlatedcreep} and
consistent with the expected geometrical phase
diagram~\cite{kolton_dep_zeroT_long}. In the context of our model, we show that
this phenomenon is related to other extremal dynamics models, particularly the
Bak-Sneppen (BS) model~\cite{bs}.

\section{Models and reference results}

A prototypical description of the depinning transition and the regimes of creep
and flow it separates is provided by the quenched Edwards Wilkinson (qEW) model.
In a one-dimensional geometry, the model can be defined by the stochastic
equation~\cite{Ferrero_CRP_2013}
\begin{equation}
\gamma\partial_t u(z,t)=c\partial_z^2u(z,t)+f+f_p(u,z)+\eta(z,t),
\label{eq:qew}
\end{equation}
representing the overdamped driven dynamics for an elastic interface, whose
position is parametrized by a univalued displacement field $u(z,t)$
with stiffness $c$ and friction constant $\gamma$. The pinning force
$f_p(u,z)=-\partial_u U(u,z)$ represents the effects of a random-bond type of
disorder described by the bounded potential $U(u,z)$, and $f$ is the uniformly
applied external driving. Thermal effects are incorporated through the white
noise correlated term, $\eta(z,t)$, with $\langle\eta(z,t)\rangle=0$ and
$\langle\eta(z,t)\eta(z',t')\rangle =2\gamma T \delta (t-t')\delta (z-z')$. In
numerical implementations the continuous spatial coordinate $z$ is typically
replaced by discrete spatial points, which are labeled by a discrete index $i$.
This model has been extensively studied. In spite of this, accurate enough
values of relevant critical exponents at $T=0$ were numerically obtained only
very recently, after acknowledging subtle power-law corrections when simulating
very large systems. The reported values are $\beta = 0.245 \pm 0.006$, $z =
1.433 \pm 0.007$, $\zeta=1.250 \pm 0.005$, and $\nu=1.333 \pm 0.007$ for the
depinning, dynamical, roughness, and correlation length
exponents~\cite{exponentes}.

Here, we will work with a particular version of this discretized model that
adapts particularly well to numerical simulations and permits us to study in a
more precise and controlled way some key properties of the thermal rounding
problem. In this implementation, the potential energy landscape $U_i(u)$ is
assumed to consist of very narrow wells located at random positions along the
$u$ direction different for every $i$. When the interface is pinned to a given
well, its location can be considered to be fixed (since the well is very
narrow). To be taken out of the well, a threshold force must be applied. This
threshold force is noted $f_i^{\text{th}}$. Different wells have different
values of threshold forces, namely $f_i^{\text{th}}$ is a stochastic variable.
The energy landscapes at different spatial positions $i$ are assumed to be
totally uncorrelated. It is expected that in the presence of a finite driving
force, and for a sufficiently dense distribution of pinning wells, each point of
the interface must necessarily sit in one potential well, namely there are no
equilibrium position in which some point of the interface is in a flat region of
the pinning potential.

The state of the system can be characterized by the set of values $f_i$ that
represents by definition the total elastic force acting on a particle trapped at
a given site $i$ [i.e., the first term on the right-hand side of
Eq.~(\ref{eq:qew})]. The state of the system is thus stable if
$f_i+f<f_i^{\text{th}}$ for every $i$. The meaning of $f_c$ in this scheme is
the following. There exist stable configurations of the system if the applied
force $f$ is lower than some critical value $f_c$. On the other hand, when
$f>f_c$ there are always sites for which $f_i+f>f_i^{\text{th}}$. The
critical force $f_c$ has a well-defined limit in the thermodynamic limit,
provided the explored sample maintains a correct aspect
ratio~\cite{bolech2004,kolton2013}. In this case, the temporal dynamics is
assumed to proceed as follows.  In a unitary time step, any site $i$ for which
$f_i+f>f_i^{\text{th}}$ moves forward a distance $\lambda$ to the next pinning
well ($\lambda$ is the separation between consecutive random pinning centers
and, thus, exponentially distributed). This produces a modification on the
elastic force $f_i$ and those of the neighbor sites $i\pm 1$ according to (a
unitary spring constant is assumed)
\begin{eqnarray}
f_i &\rightarrow& f_i-2\lambda\nonumber \\
f_{i+1} &\rightarrow& f_{i+1}+\lambda\label{fi}\\
f_{i-1} &\rightarrow& f_{i-1}+\lambda\nonumber
\end{eqnarray}
All sites that are unstable at a given time step are updated in parallel. After
the update, the new values of $f_i+f$ and $f_i^{\text{th}}$ are compared, and the new
unstable sites for the next time step are determined. The number of unstable
sites at a given time step divided by the system size determines the
instantaneous velocity of the interface~\footnote{With this definition, the
velocity tends to one as $f$ becomes very large. Usually, the most realistic
situation is that velocity reaches the so-called fast-flow regime where $v
\propto f$. The difference is due to the fact that we do not take into account
the finite time necessary to jump from one potential well to the next. This
difference, however, has no effect on quantities in the small region around
$f_c$ and small temperatures which is our main interest here.}. The average
velocity is calculated as the temporal average of this quantity.

Note that we generate new random pinning centers as the values of the coordinate
$u$ of the interface increase along the simulation, and they never repeat.
Namely, the boundary conditions are open in the $u$ direction. This is known to
lead to some spurious effects associated to very rare pinning configurations
that can pin the system for any applied external
force~\cite{bolech2004,kolton2013}. However, we found that these effects are not
relevant for our system size (in the spatial direction), for the time length of
our simulations, and for the thresholds distribution chosen.

\begin{figure}
\includegraphics[width=0.9\columnwidth,clip=true]{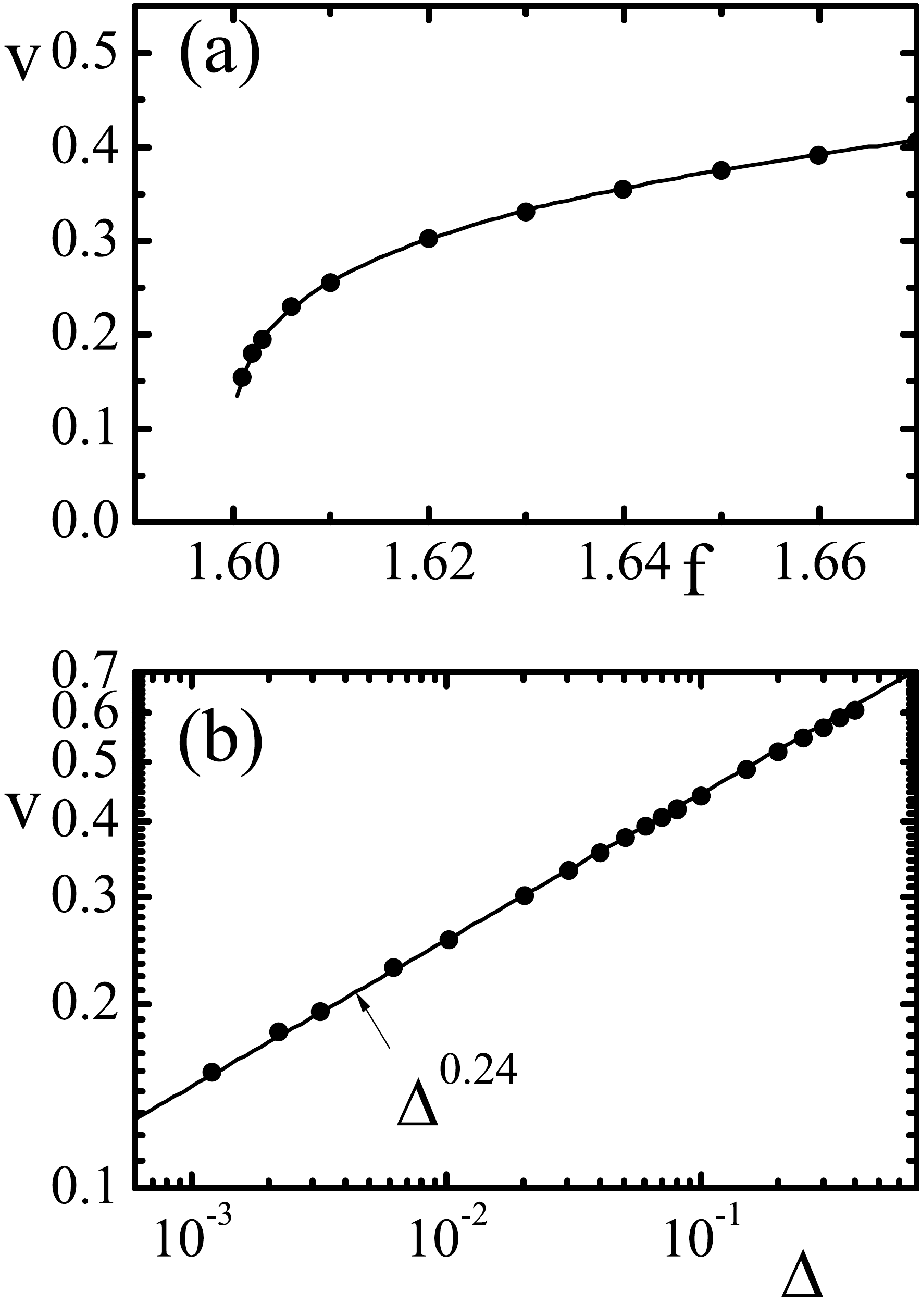}
\caption{(a) Velocity as a function of the applied force for a qEW model with $3
\times 10^4$ sites, using the narrow well form of the pinning potential.
Continuous line is a fitting using a function $v\sim (f-f_c)^\beta$. The fitted
values of $f_c$ and $\beta$ are $f_c=1.600\pm 0.001$, $\beta=0.24 \pm 0.01$. (b)
Same data in logarithmic scale, with $\Delta\equiv (f-f_c)$.}
\label{fig:beta_qew}
\end{figure}

\begin{figure}
\includegraphics[width=\columnwidth,clip=true]{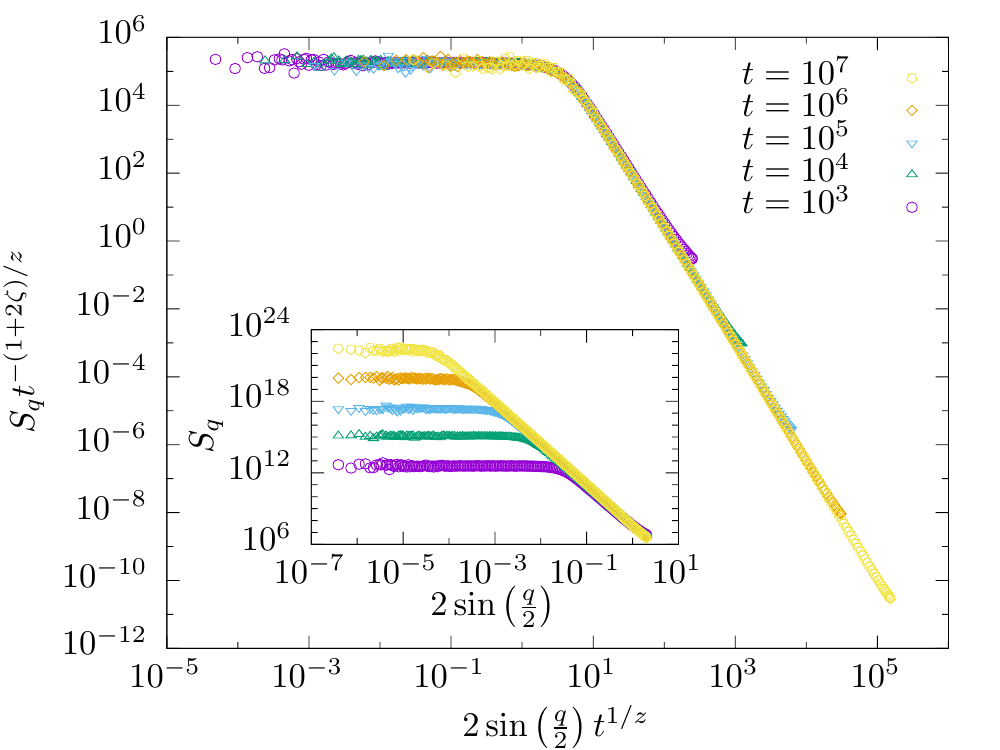}
\caption{Time-dependent structure factor $S_q(t)\equiv \langle
|u_q(t)|^2 \rangle$ for an initially flat interface at $t=0$ right at $f_c$, for
the model of Eq.~(\ref{eq:qew}) {with narrow pinning wells}. Inset: raw data for
different times. Main: scaled data using the universal non-stationary relaxation
form $S_q(t) \sim q^{-(1+2\zeta)}G(qt^{1/z})$ and the precisely
known~\cite{exponentes} dynamical and roughness depinning exponents, $z$ and
$\zeta$, respectively.}
\label{fig:sq_scaled}
\end{figure}

First we will provide numerical evidence that this simple scheme produces
results that are not only qualitatively equivalent to those obtained with the
continuous model of Eq.~(\ref{eq:qew}) but also quantitatively correct
regarding $T=0$ universal depinning exponents. We numerically calculated the
velocity $v$ as a function of the applied force $f$ for a system of $3\times
10^4$ sites, with an average separation between wells of $1/2$ (i.e., $\overline
\lambda=1/2$), and a distribution of $f^{\text{th}}$ given by a Gaussian with unitary
mean and unitary variance~\footnote{This produces also negative values of
$f^{\text{th}}$, that seem unphysical. However, note that the mean value of $f^{\text{th}}$ is
irrelevant, and could be shifted to arbitrarily larger values (with the
corresponding shift of all forces in the system), in such a way that negative
values of $f_{\text{th}}$ do not appear.}. The results are  presented in
Fig.~\ref{fig:beta_qew}. They were fitted with a power law of the form $v\sim
(f-f_c)^\beta$, where the values of $f_c$ and $\beta$ were freely adjusted. The
best fitting provides $f_c=1.600\pm 0.001$ and $\beta=0.24\pm 0.01$. The value
of $\beta$ coincides very well with the best value determined in the continuous
potential version of the model~\cite{exponentes}, giving confidence that the
narrow potential well approximation does not introduce qualitative changes in
the properties of the depinning transition. Furthermore, in
Fig.~\ref{fig:sq_scaled} we show that the structure factor $S_q(t)\equiv
\langle |u_q(t)|^2 \rangle$ of the configuration $u_i(t)$ generated by our model
as a function of the discrete time $t$, displays an excellent agreement with
the nonstationary scaling $S_q(t) \sim q^{-(1+2\zeta)}G(q t^{1/z})$, predicted
for the relaxation of an initially flat interface, $u_i(0)=0$, by using the
roughness and dynamical depinning exponents $\zeta=1.25$ and $z=1.43$ accurately
obtained for the continuous qEW model right at $f_c$~\cite{exponentes}. Note
that the identity $\beta=(z-\zeta)/(2-\zeta)$, expected from the statistical
tilt symmetry of the qEW model~\cite{Ferrero_CRP_2013}, is well satisfied. We
conclude that at $T=0$ both the stationary and the nonstationary relaxation
dynamics near the depinning transition is well described by the qEW depinning
exponents. Our simplified model, at $T=0$, thus belongs to the qEW universality
class.

\section{Thermal rounding of the depinning transition}

We now discuss the way in which a fluctuating external field rounds the
depinning transition. We start, in Sec.~\ref{sec:uniform}, with the
simpler case of the rounding effect of an uniform activation rate for trapped
particles. Afterwards, we consider the effect of thermal fluctuations in
Sec.~\ref{sec:arrhenius}.
In both cases we have considered that each activation let a particle jump only
to the next trap, in the forward direction. This approximation is justified near
the depinning transition at small temperatures. Due to the driving force, a
particle requires much more energy to reach the trap behind it than to reach the
trap in front of it: if a particle has to jump a barrier $U$ to escape from its
trap, to reach the trap behind it implies to overcome a larger finite barrier,
$U' \sim U + f  \lambda$, where $\lambda$ is the typical distance between
traps. Note that this remains true even if $U \to 0$, when the particle is near
destabilizing. Hence, at small but finite $T$, the backwards jump is not only
exponentially prohibitive, but also very improbable in a purely diffusive regime
(no barriers at all) due to the forward finite drive $f \sim f_c$.

\subsection{Uniform activation rates}
\label{sec:uniform}

We first consider the effect of an ``uniform activation rate'' on the depinning
transition. This peculiar but still stochastic force is meant to activate sites
with a fixed probability $h$ at every time step and spatial position in the
system, independent of the interface state. In the context of the interface, we
can think $h$ as a uniform activation for all particles with ${(f_i + f)} <
f^{\text{th}}_i$, regardless of the actual values of $f_i$, in sharp contrast with the
Arrhenius case where the activation does depend on $f_i$ values through the
energy barrier to escape a trap. That is, we will consider that any pinned
particle can escape its trap with the same probability $h$. In the following, we
will indistinctly call $h$ the ``activation field'' or rate, or simply the
``external field,'' to emphasize the analogy with the spin model. On the other
hand, particles with $f_i > f^{\text{th}}_i$ will always be activated.

It is clear that the velocity of the interface (i.e., the number of active sites
per unit time) will be larger in the presence of a finite $h$ than for $h=0$.
For small values of $h$ and close to the critical force, the velocity is
expected to follow a scaling relation as in Eq.~(\ref{eq:v_t_f}), namely
\begin{equation}
v(\Delta,h) \sim h^{\psi_h} g\left (\Delta/h^{{\psi_h}/\beta}\right ).
\label{eq:scaling}
\end{equation}
We use $\psi_h$ for the rounding exponent in this case, to emphasize the fact
that we are applying a uniform activation probability $h$. Remarkably, in this
case the value of $\psi_h$ can be obtained in terms of the other exponents of
the transition. The argument leading to this conclusion is contained
in Ref.~\cite{hinrichsen}, p.~47, for the directed percolation (DP) transition.
Here we present the argument in a slightly different form, for the rounding of
the qEW depinning transition. For $\Delta<0$ the value of $v$ would be zero were
it not for the existence of a finite $h$. The finite $h$ triggers a number of
avalanches with a density that is (for small $h$) simply proportional to $h$.
The average size of each of those avalanches (its ``mass'' in the language
of Ref.~\cite{hinrichsen}) is a property of the model in the $h=0$ limit and
is given by $\sim |\Delta| ^{-(d+z)\nu+2\beta}$ (allowing for an arbitrary
spatial dimension $d$). It is thus obtained that
\begin{equation}
v \sim h |\Delta|^{-(d+z)\nu+2\beta}
\label{eq:v_from_dp}
\end{equation}
for $\Delta<0$. The condition for this equation to be valid is that $h$ is so
small that two different activated clusters do not overlap.

The condition that Eq.~(\ref{eq:v_from_dp}) is compatible with
Eq.~(\ref{eq:scaling}) means that the function $g(x)$ must behave for large $x$
as $g(x)\sim x ^{-(d+z)\nu+2\beta}$, leading to
\begin{equation}
\psi_h= \frac{\beta}{(d+z)\nu-\beta}.
\label{eq:psi}
\end{equation}

\begin{figure}
\includegraphics[width=0.9\columnwidth,clip=true]{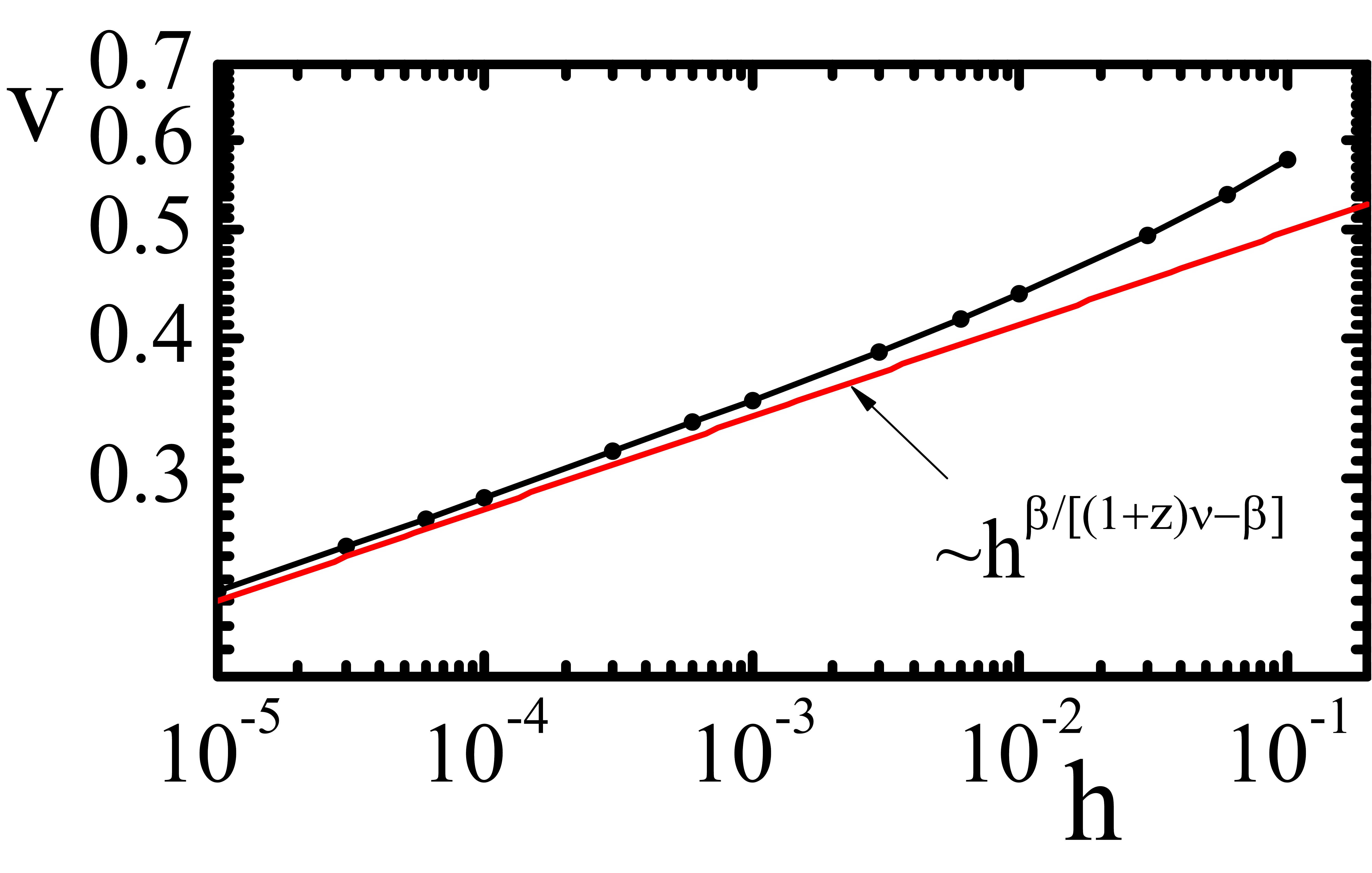}
\caption{Velocity $v$ as a function of the intensity of the
external field $h$, at the critical point. Points are the results of the
simulation. Straight red lines is drawn with the expected slope $\psi_h\simeq
0.082$. The numerical data approach the theoretical prediction for sufficiently
low values of $h$.}
\label{fig:psitilde}
\end{figure}

The numerical value of $\psi_h$ predicted by Eq.~(\ref{eq:psi}) (using the best
known values of $\beta$, $z$, and $\nu$, for qEW in $d=1$) is $\psi_h\simeq
0.082$. Results of numerical simulations of our implementation of the model
right at the critical point produce the results in Fig.~\ref{fig:psitilde}. The
results of the numerical simulations are consistent with the analytical
prediction as $h$ is reduced. In fact, this is the limit in which the analytical
prediction was obtained.

\begin{figure}
\includegraphics[width=0.9\columnwidth,clip=true]{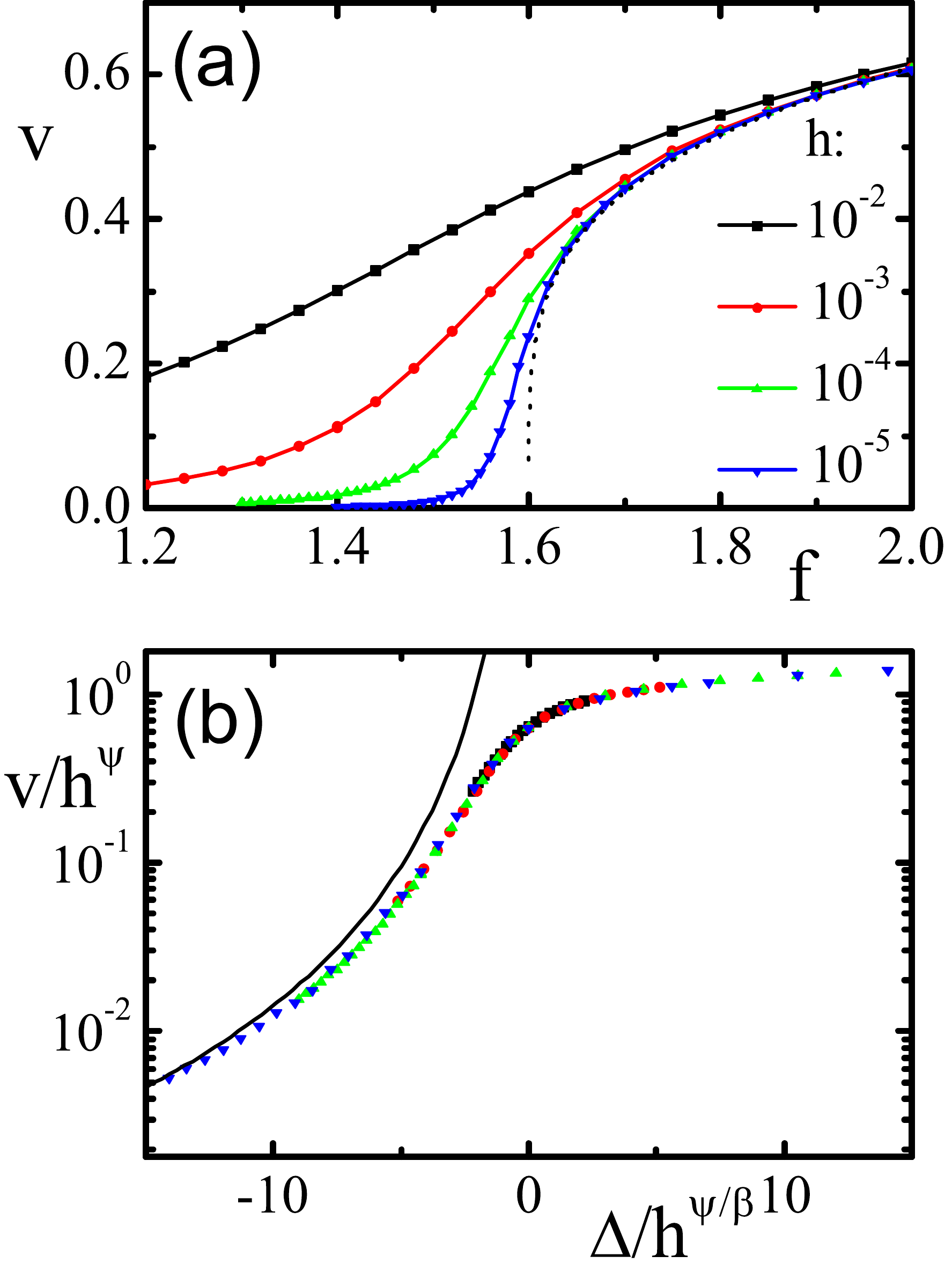}
\caption{(a) Velocity as a function of external field for qEW
model, below and above the critical point, for different values of $h$, as
indicated. The $h=0$ critical behavior is indicated by the dotted line. (b) Same
results, rescaled using the expected values of $\beta=0.245$ and $\psi_h=0.082$
($\Delta=f-f_c$). The continuous line is the asymptotic form predicted by
Eq.~(\ref{eq:v_from_dp}).}
\label{fig:psitilde_scaling}
\end{figure}

We will now show that the full scaling law Eq.~(\ref{eq:scaling}) is satisfied.
Figure~\ref{fig:psitilde_scaling}(a) shows results of simulations at finite
value of $h$, both below and above criticality. In
Fig.~\ref{fig:psitilde_scaling}(b) we show how all curves can be scaled onto a
unique universal curve using the appropriate value for the exponents $\beta$ and
$\psi_h$. In particular, for large and negative $\Delta$, the data converge to
the expected behavior from Eq.~(\ref{eq:v_from_dp})~\cite{hinrichsen}. We thus
verify the validity of Eq.~(\ref{eq:psi}) for qEW under uniform activation
rates.

\begin{figure}
\includegraphics[width=0.9\columnwidth,clip=true]{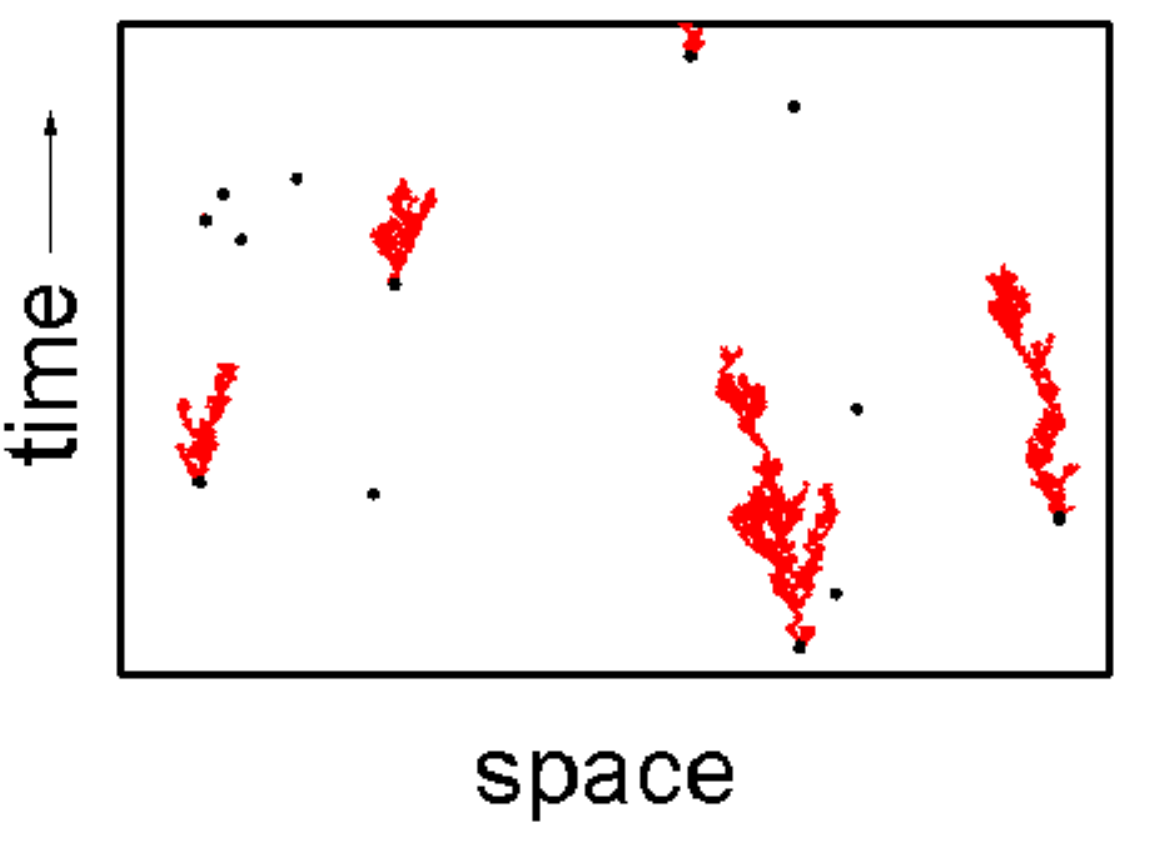}
\caption{Space and time distribution of active sites in the
system, for the qEW case. Small (red) dots are sites that are activated by
neighbor sites in the previous time step. Larger (black) dots are sites
activated at the uniform rate $h$, and thus randomly distributed. The plot was
obtained below criticality, at $f=0.54$, $h=10^{-3}$. Note the structure of
independent clusters, each of them initiated by a site activated by the field.
The space-time span of the graph is 1000 sites and 1500 time steps.}
\label{fig:activity_below_fc}
\end{figure}

\begin{figure}
\includegraphics[width=0.9\columnwidth,clip=true]{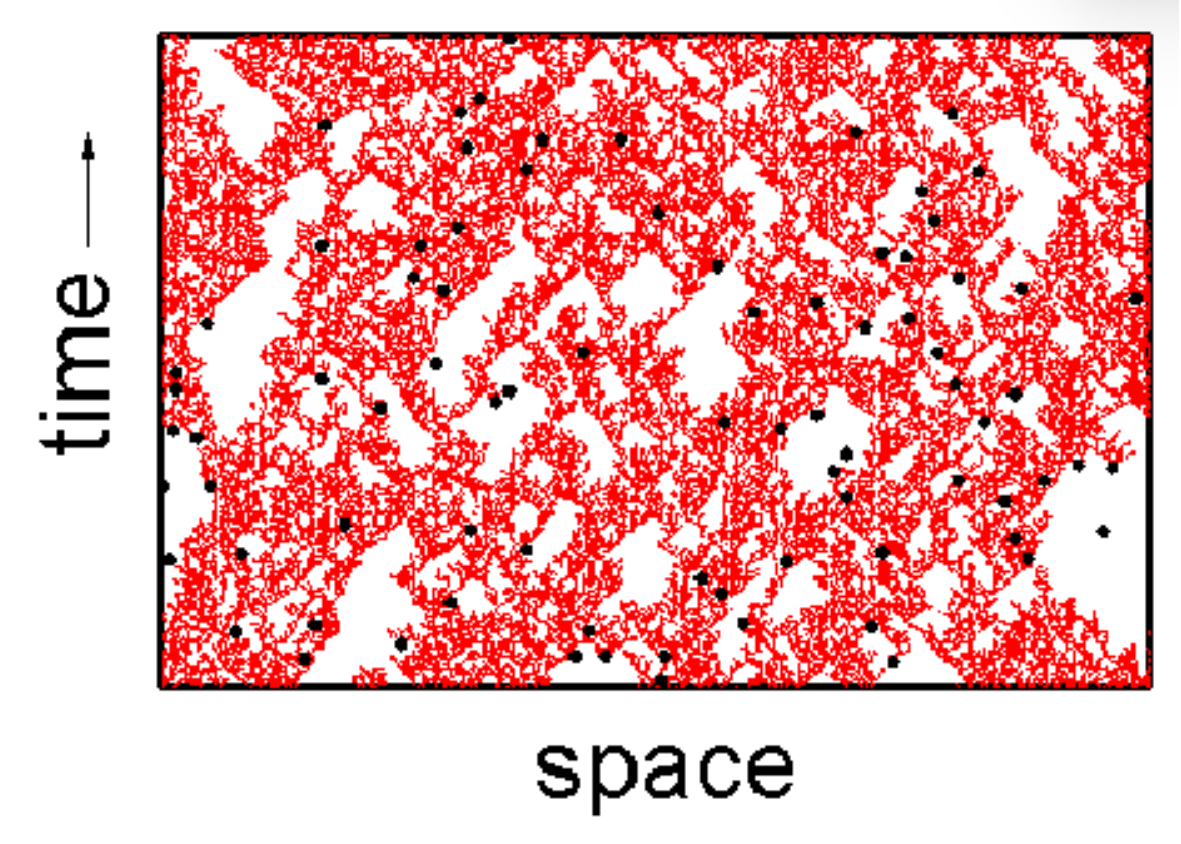}
\caption{Same as previous figure, at $f=f_c\simeq 1.600$,
$h=10^{-3}$. The span of the graph is 300 sites and 450 time steps.}
\label{fig:activity_at_fc}
\end{figure}

It is instructive to observe the activity pattern in the system, in the presence
of the external field $h$ controlling the uniform activation rate. In
Figs.~\ref{fig:activity_below_fc} and \ref{fig:activity_at_fc} we show
spatiotemporal plots of the activity, where active sites are indicated. Active
sites are separated in two classes: standard active sites (which are activated
by other active sites in the previous time step) and sites activated directly
by the field $h$. Below the critical point, the structure of activity follows
the trend that was assumed  in deriving Eq.~(\ref{eq:v_from_dp}), namely a
sparse and uncorrelated set of sites activated by the field, each of them
generating a cluster of activity. Increasing the values of $f$, or $h$
(Fig.~\ref{fig:activity_at_fc}), we see how the activity percolates across the
system. It is worth noting here that the so-called ``mass'' $|\Delta|^{-\nu
(d+z)+2\beta}$, measuring the amount of active sites per unit time in one
avalanche and giving place to Eq.~(\ref{eq:v_from_dp}), is different from the
quantity $|\Delta|^{-\nu(d+z)}$ naively expected using the characteristic
spatial size $|\Delta|^{-\nu d}$ and the characteristic time $|\Delta|^{-\nu z}$
of a depinning-like avalanche~\cite{kolton_dep_zeroT_long}. This is so because
avalanches of activity are actually porous objects in the ($d+1$-dimensional)
space time. This porosity can be appreciated in
Figs.~\ref{fig:activity_below_fc} and \ref{fig:activity_at_fc}.

Scaling relation Eq.~(\ref{eq:psi}) is remarkably similar to the hyper scaling
relation in equilibrium statistical mechanics. In fact, the order parameter of
an equilibrium second-order phase transition right at the critical temperature
vanishes as a function of the activation field $h$ as $\sim h^{1/\delta}$, in
terms of the so-called magnetic exponent $\delta$, which obeys the scaling
relation \cite{Stanley71}
\begin{equation}
\delta^{-1}= \frac{\beta}{d\nu-\beta}.
\label{eq:hyper_scaling}
\end{equation}
This follows from the generalized homogeneity property of the free energy close
to the critical point. Since the meaning of $\delta^{-1}$ is formally analogous
to the thermal rounding exponent, Eq.~(\ref{eq:hyper_scaling}) has been used,
without modification, to compare with the velocity-force characteristics
obtained by numerical simulations of the driven random field Ising model
in Ref.~\cite{roters}. Here we have shown, however, that Eq.~(\ref{eq:psi}) can
be considered to be equivalent to Eq.~(\ref{eq:hyper_scaling}) if we recognize
that in the DP or qEW problem the ``time'' dimension  (that scales with an
additional factor $z$ with respect to spatial dimensions) must be added to the
internal, spatial dimension $d$. Moreover, such relation holds for $\psi_h$,
which corresponds to the particular case of uniform activation rates
(nonuniform Arrhenius activation rates are discussed in the next section). We
notice, however, that despite the formal similarity between Eqs.~(\ref{eq:psi})
and (\ref{eq:hyper_scaling}), the two expressions are obtained by very
different kinds of reasoning, the second being applied to thermal equilibrium,
while the first is applied to a far-from-equilibrium critical phenomenon.

\subsection{Arrhenius activation rates}
\label{sec:arrhenius}

\begin{figure}
\includegraphics[width=\columnwidth,clip=true]{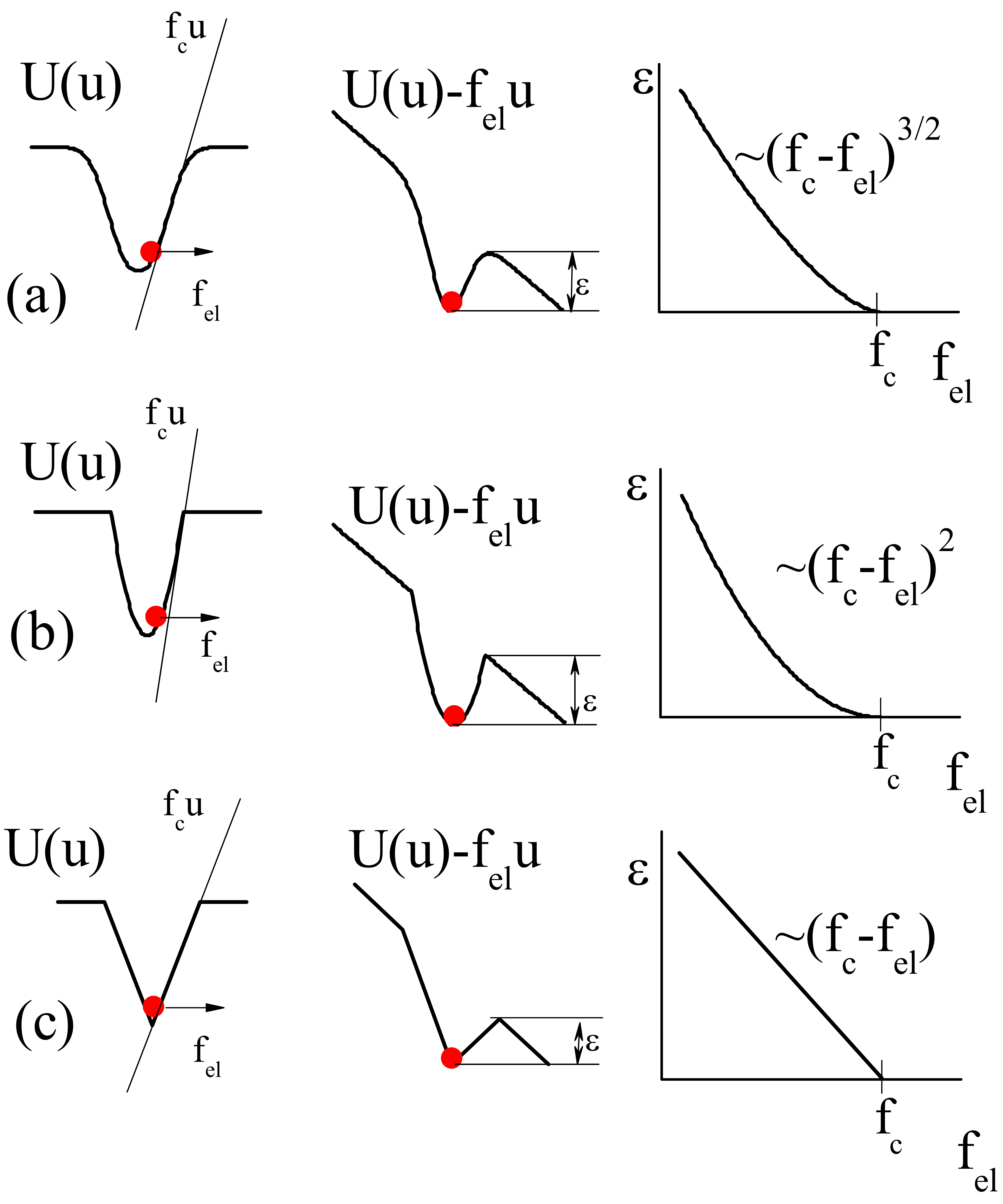}
\caption{Energy barrier as a function of the applied elastic
force $f_{\text{el}}$ for different forms of the pinning wells. (a) Smooth
pinning case. (b) Sharp-ending parabolic pinning well. (c) Triangular well. The
energy barrier $\epsilon$ behaves as $\epsilon\sim(f_c-f_{\text{el}})^\alpha$,
with $\alpha=3/2$, $\alpha=2$, and $\alpha=1$ respectively. Note that for one
particle $f_c \equiv f_i^{\text{th}}$.}
\label{fig:sketch}
\end{figure}

The fundamental difference between thermal activation and the uniform activation
discussed in the previous section is that for the latter the activation
probability is the same at each time step for each nonactive site, whereas the
effect of temperature depends on the value of an energy barrier that has to be
overcome. In the narrow well approximation of the qEW case, given a stable
configuration of the system, energy barriers can be naturally identified in
terms of the elastic forces $f_i$ and maximum threshold forces
$f^{\text{th}}_i$. For convenience we define $x_i=f^{\text{th}}_i-f_i$. The
energy barrier for site $i$ (noted $\epsilon_i$) vanishes when the applied force
$f$ is such that $f\to x_i$. Typically, $\epsilon_i \sim (x_i-f)^\alpha$. The
value of $\alpha$ depends on the shape of the pinning potential (see
Fig.~\ref{fig:sketch}). For the most standard case in which the pinning
potential is smooth (more precisely, with a continuous second derivative), the
value of $\alpha$ is $3/2$. On the other hand, if the pinning potential is as
depicted in Fig.~\ref{fig:sketch}(b) or \ref{fig:sketch}(c), $\alpha=2$ or
$\alpha=1$ is, respectively, obtained. In any case, the value of $\alpha$ is
well defined once the characteristics of the pinning potential are defined (see
Appendix~\ref{sec:single_par} for a general discussion of one particle dynamics
in a periodic potential with an anomalous marginality at the critical force).

The effect of temperature can be incorporated in the dynamical algorithm of the
previous section in the following way. At each time step the values of $x_i$ are
calculated. Those sites with $x_i<f$ are automatically active, as in the $T=0$
case. Sites with $x_i>f$ are activated with some probability $p_i$ according to
an Arrhenius law, namely
$p_i=\exp(-\epsilon_i/T)=\exp(-(x_i-f)^\alpha/T)$~\footnote{The Arrhenius
activation formula is a valid approximation whenever the local barrier $(x_i-
f)^\alpha$ is larger than the temperature. This condition may strictly fail for
particles near an instability, where $x_i \sim f$. In the small temperatures
thermal rounding regime, however, this situation represents a very narrow layer
of $x_i \sim 0$. We have checked that corrections to the Arrhenius activation
formula taking care of these rare cases does not affect our conclusions.}.
A wide distribution of barriers leads, in general, to a power-law distribution
of activation times~\cite{jstat,vinokur}. In turn, this implies a nontrivial
correlation of the activation process: the probability that a site becomes
active in the next time interval $\tau$, given that we know it has not been
active for some time interval $\tau_0$ decreases as $\tau_0$ increases; i.e.,
the process acquires an effective ``memory.'' This is in contrast to what
happens for a constant activation probability $h$ since in that case the forward
in time probability is independent of the previous history. This simple
qualitative idea can be used to understand the activation pattern of a spatially
extended system. If in some spatial region there has not been activity for some
time, it means that the values of the energy barriers $\epsilon$ in that region
are rather large, and then it is likely that those sites will not be thermally
activated soon. On the other hand, in regions in which there are active sites,
it is likely that some of them will fall in potential wells with small values of
$\epsilon$, and thus will activate again soon, maintaining the activity in that
region.

\begin{figure}
\includegraphics[width=0.9\columnwidth,clip=true]{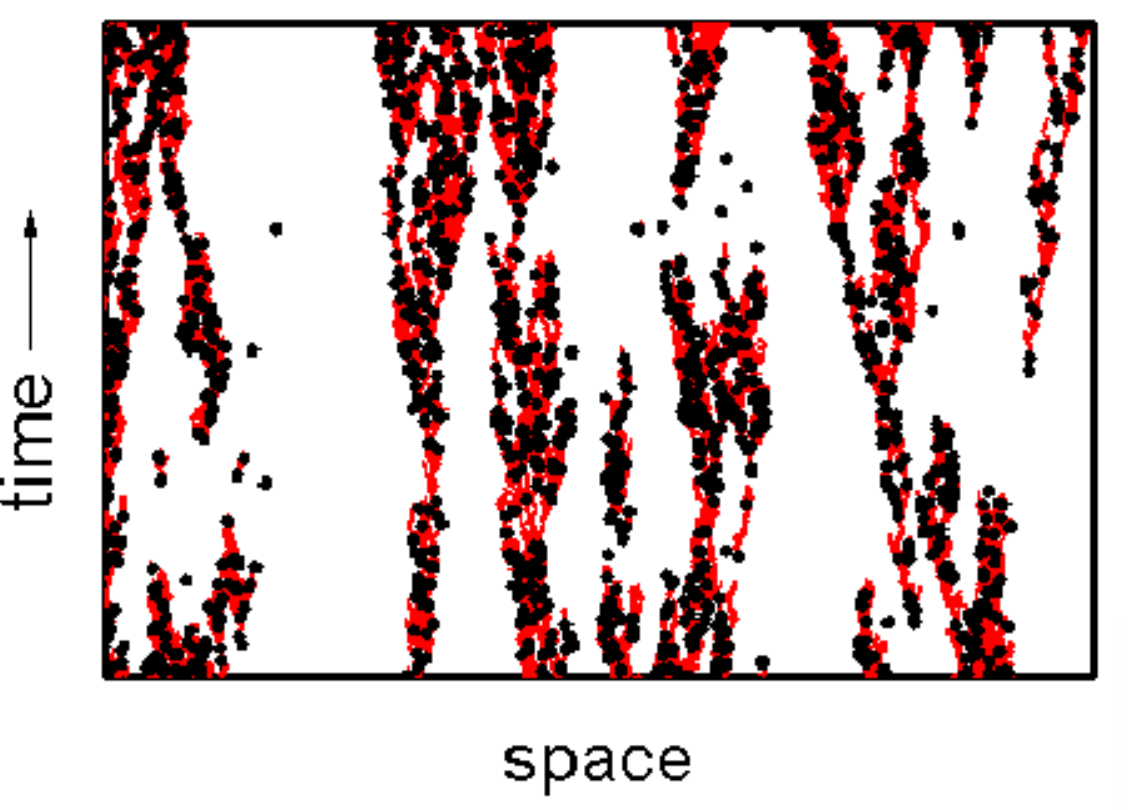}
\caption{Same as Fig.~\ref{fig:activity_at_fc} but for a finite
temperature ($\alpha=1$), instead of an uniform activation rate. Parameters of
the simulation are $f=0.5$, $T=0.01$. Note how now the location of thermally
activated sites is strongly correlated spatially and temporally. The span of the
graph is 1500 sites and 1000 time steps.}
\label{fig:qew_activados_contemp}
\end{figure}

To illustrate this, we show in Fig.~\ref{fig:qew_activados_contemp} results
equivalent to those in Fig.~\ref{fig:activity_at_fc}, but for the case of a
finite temperature instead of a constant and uniform activation rate. The
nonuniform spatial and temporal distribution of thermally activated sites is
apparent in this figure, making clear that the effect of temperature is much
more subtle than that of a uniform activation field~\cite{jstat}.

\begin{figure}
\includegraphics[width=0.9\columnwidth,clip=true]{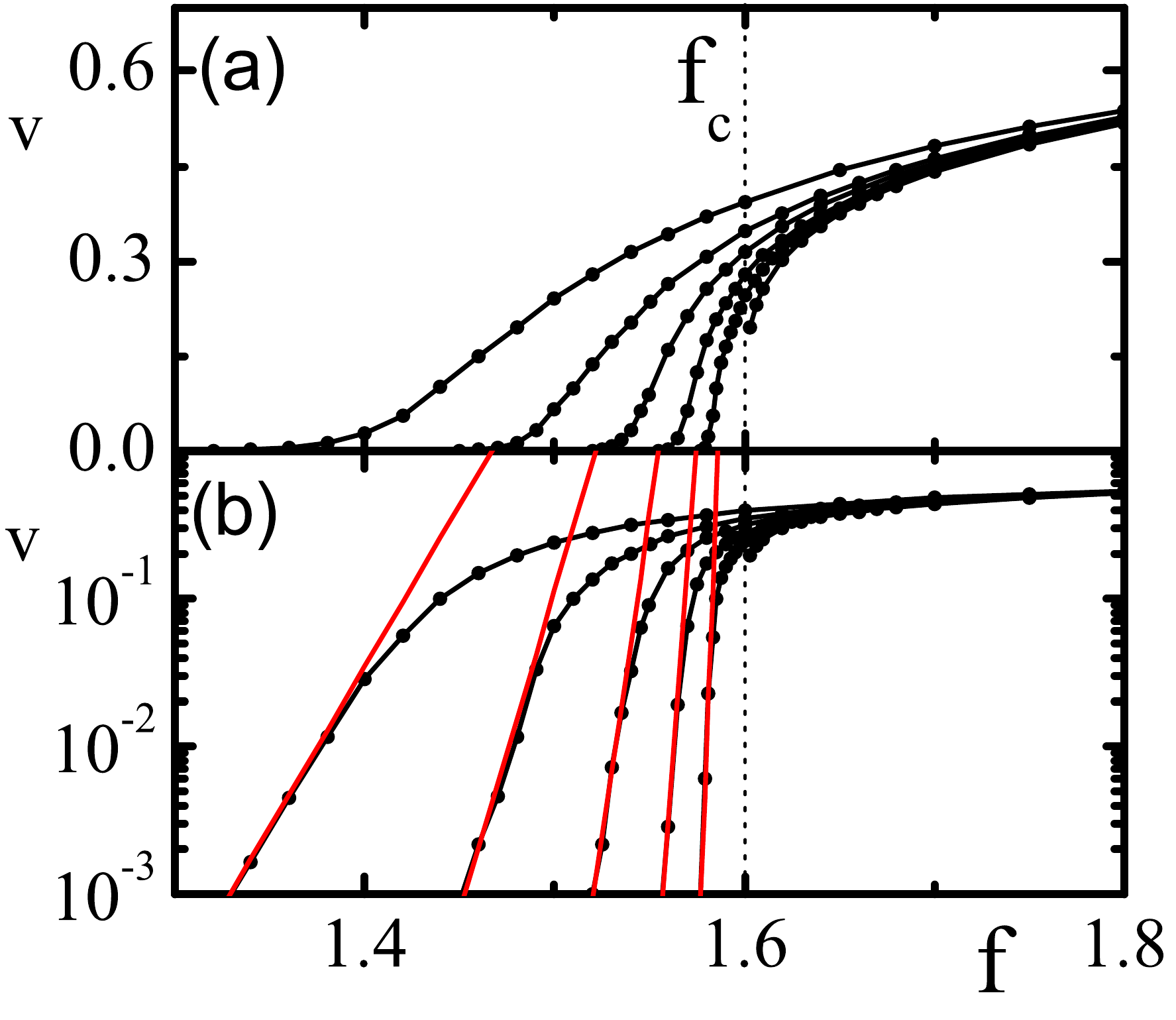}
\caption{Velocity as a function of applied force, at different
temperatures in a system of $10^4$ sites with $\alpha=1$. Temperature is
$T=0.02$ for the left-most curve, and is divided by two for each successive
curve linear (a) and logarithmic (b) scale. In (b), the red straight lines
display the form $v=C T^{q} \exp(\Delta/T)$, with $q=1.7$ and a single $C$
factor for all curves. For a justification of the $T^q$ factor see
Sec.~\ref{sec:depinning}.}
\label{fig:th0}
\end{figure}

We focus now on the effect of temperature on the average velocity of the
interface. Figure~\ref{fig:th0} shows results of velocity as a function of
applied force at different temperatures, with the $\alpha$ exponent of the
pinning wells equal to $1$, and using the standard thermal activation algorithm
previously described, namely: sites are active (and then jump to the next
potential well) either with probability $1$, if $x_i<f$, or with probability
$\exp (-(x_i-f)^{\alpha}/T)$, if $x_i>f$.  Although at first sight the results
in Fig.~\ref{fig:th0}(a) look similar to those in
Fig.~\ref{fig:activity_at_fc}(a), one important difference is the fact that at
low values of $f$, an activated dependence of the form
\begin{equation}
v\sim \exp( -|\Delta| /T)
\label{eq:activation_barriers}
\end{equation}
is observed, as shown by the red lines in Fig.~\ref{fig:th0}(b). This
exponential dependence can be naturally interpreted as indicating the existence
of activation barriers of height $|\Delta|$ in the system. In fact, the assumed
existence of such barriers led Middleton~\cite{middleton} to suggest that for
thermal rounding, the compatibility between Eqs.~(\ref{eq:activation_barriers}),
(\ref{eq:v_t_f}), and (\ref{eq:v_f_beta}) implies that
\begin{equation}
v(\Delta,T)\sim T^\beta g\left (\Delta/T\right)
\label{eq:psi_beta}
\end{equation}
namely, the prediction $\psi=\beta$ is obtained~\footnote{Note that in the case
 of an $ \alpha$ value different from one, the prediction becomes
$\psi=\beta/\alpha$, which is the exact result expected for the depinning or
saddle-node bifurcation with a normal form $\dot{x}=x^\alpha+\epsilon$ of a
particle in a periodic potential (see Appendix~\ref{sec:single_par}).}.

\begin{figure}
\includegraphics[width=0.9\columnwidth,clip=true]{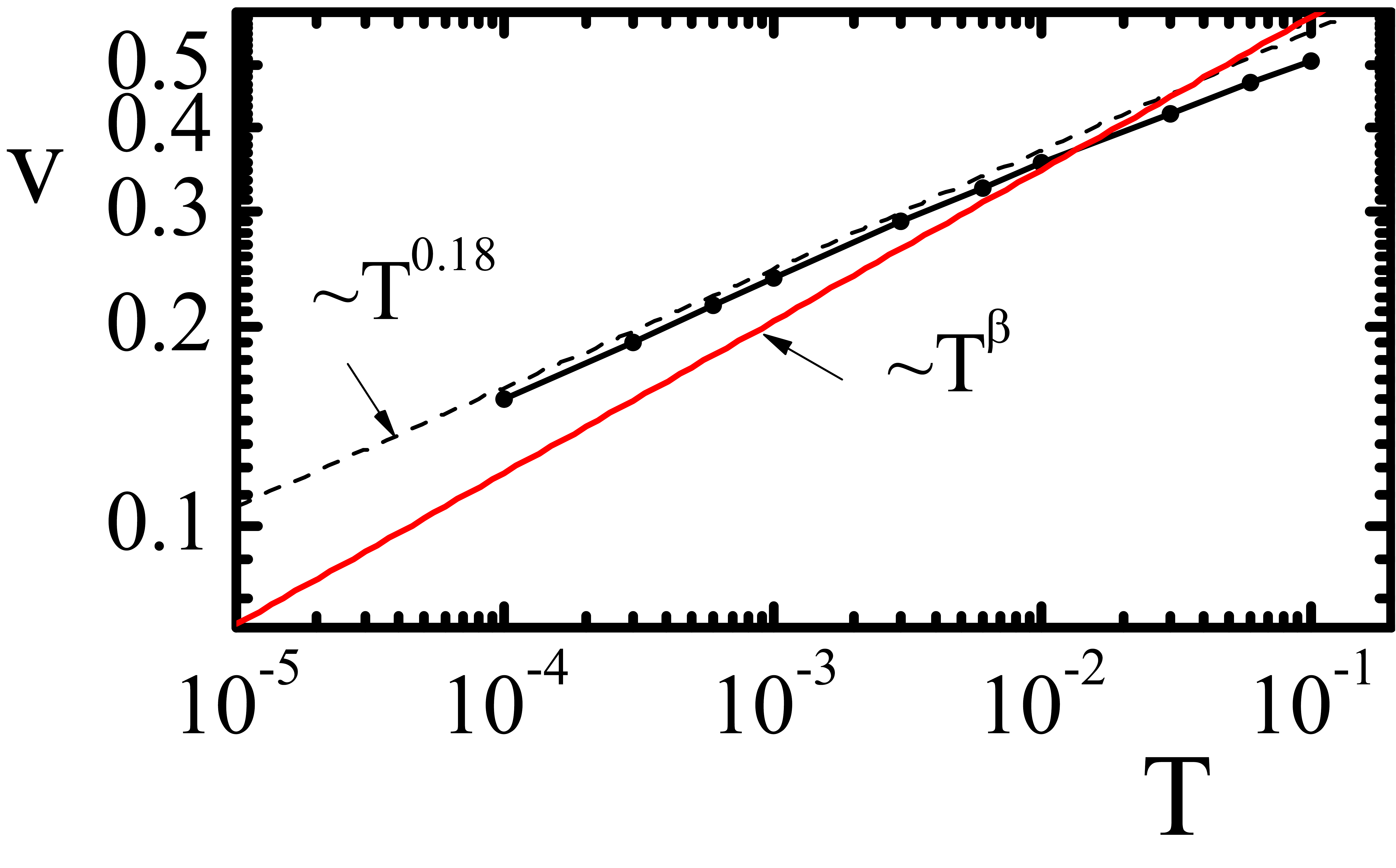}
\caption{Velocity $v$ as a function of temperature at the
critical point. The red straight line has a slope $\beta\equiv 0.245$, which is
the value predicted by the arguments of Middleton~\protect\cite{middleton}. The
numerical results seem to point to a value $\simeq 0.18$, instead (dotted
line).}
\label{fig:psitilde2}
\end{figure}

\begin{figure}
\includegraphics[width=0.9\columnwidth,clip=true]{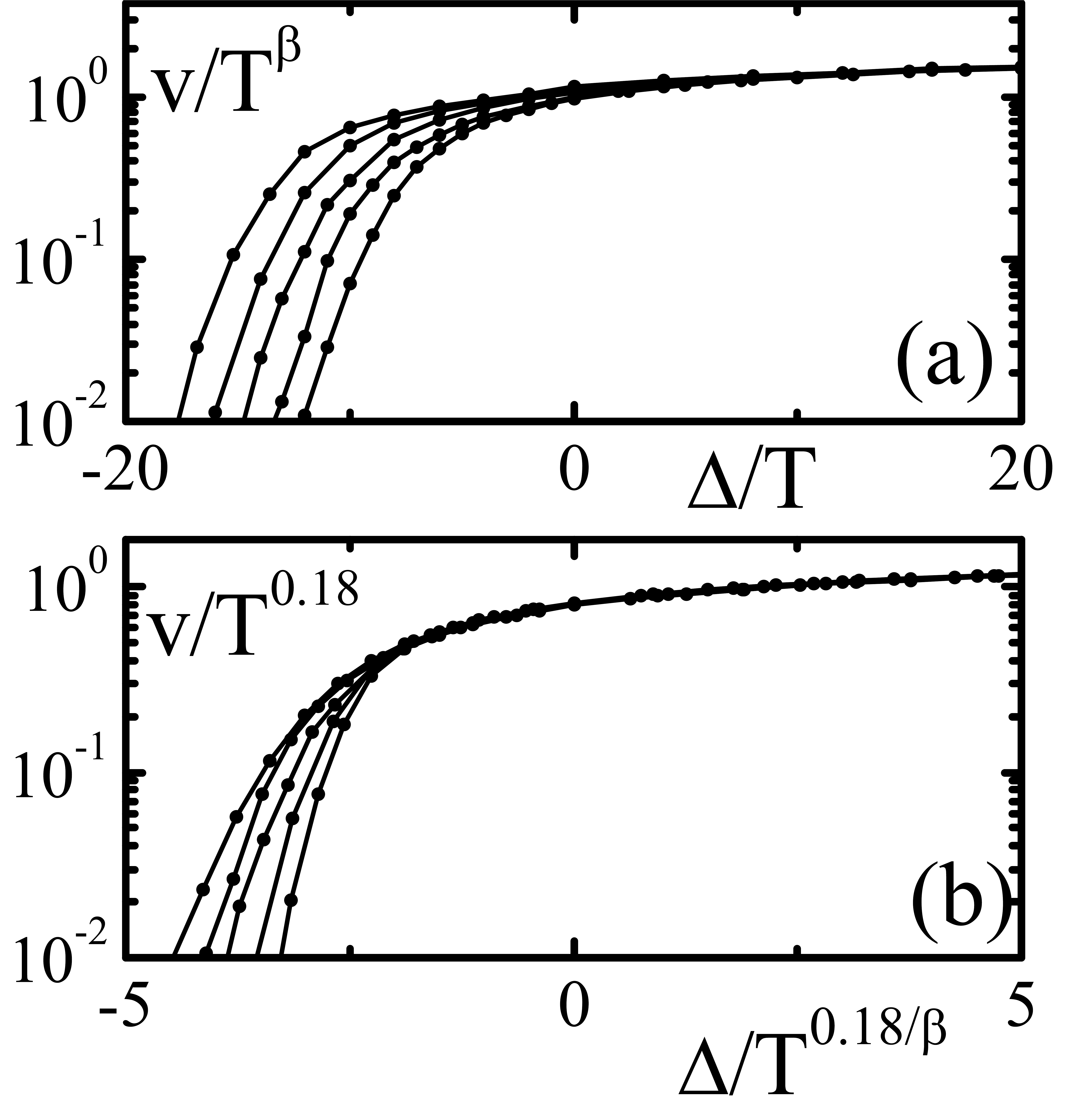}
\caption{The data in Fig.~\ref{fig:th0} scaled according to (a) the Middleton
suggestion, and (b) using the value 0.18 for the thermal rounding exponent,
according to results in Fig.~\ref{fig:psitilde2}. None of the cases permits a
correct scaling of the whole data.}
\label{fig:otrosc}
\end{figure}

To verify the prediction $\psi=\beta$, a direct run at $f=f_c$ was
performed in our model, and the results are shown in Fig.~\ref{fig:psitilde2}.
We see that the thermal rounding exponent obtained seems to be around 0.18,
clearly below the expected value of $\psi=\beta\simeq 0.245$. The discrepancy
appears also quite clearly when trying to scale the set of curves $v(\Delta,T)$,
according to the Middleton scaling in Eq.~(\ref{eq:psi_beta}). This is done in
Fig.~\ref{fig:otrosc}(a), where we see that there are systematic deviations to
this scaling. However, with an alternative point of view, we may argue that the
result in Fig.~\ref{fig:psitilde2} must be taken into account, and the data
should be plotted according to Eq.~(\ref{eq:v_t_f}), with $\psi\simeq 0.18$.
This is done in Fig.~\ref{fig:otrosc}(b). A reasonable collapse of the curves is
observed for $\Delta>0$ and for a narrow range of $\Delta<0$, but the
exponential tail is not correctly scaled. One could thus argue that
Eq.~(\ref{eq:v_t_f}) is only valid in such reduced range. This is, however, in
sharp contrast to what was found for uniform activation rates in
Fig.~\ref{fig:psitilde_scaling}, or for the particle in a periodic potential
(see Fig.~\ref{fig:onepart} in the Appendix), where the collapse is excellent
for absolute values of the scaling variable $\Delta/T^{\beta/\psi}$ as large as
$10$. The plain conclusion to be drawn from here is that the data from the
simulations cannot be scaled according to Eq.~(\ref{eq:v_t_f}) for any value of
$\psi$.

We have found, however, that there is a way to appropriately scale the numerical
data to make them collapse onto a universal curve. Before discussing this
scaling, we shall first present a more detailed analysis of the dynamical
behavior of the system in the limit of very low temperature.

\section{Activated dynamics near the depinning threshold}

We consider now the dynamics of the system for $f<f_c$, and for vanishingly
small temperature. This is known as the creep regime~\footnote{Sometimes the
``creep regime'' is referred to the $f \to 0$ response. Here we use it to
describe the whole thermally activated flow regime below the depinning
threshold, ${f<f_c}$.}.
For a finite system, if temperature is much smaller than the other energy scales
in the problem, it can be shown~\cite{kolton_dep_zeroT_long} that the
steady-state activated dynamics of an elastic interface in a random medium
becomes essentially deterministic, with the system visiting a unique sequence of
metastable states connected exclusively by forward collective moves. To
reproduce this sequence, it is necessary to find the optimal path connecting
consecutive metastable states. The optimal path is the one that takes the system
from one metastable state to another metastable state with lower energy
overcoming the minimal energy barrier. Finding this path is a complex task even
in a discrete system, since elementary moves might consist in the simultaneous
motion of more than one particle and the enumeration of configurations grows
exponentially. Although the exact transition pathway problem can be still
approximated retaining collective moves~\cite{ferrero_correlatedcreep} using
transfer matrix techniques, in our model we can further simplify the
construction of the metastable states. If we consider only the case in which $f$
is away from zero, and sufficiently close to $f_c$, the optimal nucleus we need
to activate to move forward reduces to one particle if the disorder is strong
enough~\cite{kolton_dep_zeroT_long}. In this limit, elementary moves consist in
one-particle Arrhenius activated jumps if the temperature is still much smaller
than the local barriers $(x_i-f)^\alpha$. Backward motion is futile and can be
neglected (as we are already doing in our simulations) as it involves an extra
energy penalty of the order of $f_c \lambda$ where $\lambda$ is the typical
distance between the narrow wells. Moreover, near $f_c$ a small event may
produce in general a large forward deterministic avalanche, similar to a
depinning avalanche above threshold of a transverse size
$|\Delta|^{-\nu}$~\cite{kolton_dep_zeroT_long,ferrero_correlatedcreep}, so the
motion is mostly irreversible in this regime.

Therefore, given a pinned configuration with all $x_i>f$, thermal activation at
vanishingly small temperature proceeds by acting on the site with the smallest
activation barrier. The thermal creep regime is simpler to implement numerically
than the full thermal activation case. As the activation barrier is a monotonous
function of $x_i$, vanishing as $x_i\to f$ from above, the first site to be
activated in the creep regime is that with the smallest $x_i$. In this limit,
our model becomes identical to the so-called Zaitsev
model~\cite{Zaitsev_1992,Paczuski_1996}. This simple prescription greatly speeds
up the numerical algorithm.  So we implement thermal creep by triggering the
next avalanche precisely at the site with the smallest $x_i$. Beyond this
difference in the activation step, the developing of avalanches remains as
previously described.

\subsection{Correlated events and extremal dynamics}

Choosing the lowest $x_i$ generates a systematic increase of the values of $x_i$
over time, which brings interesting consequences into the dynamics. This is an
example of what is known as an extremal dynamics. One case in which this effect
was studied in all detail is the Bak-Sneppen (BS) model~\cite{bs}. In this
model, a random number $x_i$
between 0 and 1 is defined for each site of a 1D lattice. At each time step, the
lowest $x_i$ is selected, and is replaced by a new random number. The random
numbers corresponding to the two neighbors $i+1$ and $i-1$ are renewed too. The
systematic choice of the lowest $x_i$ produces that in the long run, a critical
value $x^*\simeq 0.6670$ builds up, such that all $x_i$ tend to be above $x^*$.
The system ``self-organizes'' in a critical state, without tuning any parameter.
This is a prototypical case of self-organized criticality.

Qualitatively, the same occurs in the qEW model in the thermal creep case.
Independently of the value of $f<f_c$, the values of $x_i$ tend in the long run
to accumulate above a value $x^*$, and a ``gap'' appears in the range $(f,x^*)$.
We will see that $x^*=f_c$, implying that eventually the configuration of the
system becomes a critical depinning one. This will allow us to explain, through
a much simpler model, the results previously obtained with the exact transition
pathway algorithm~\cite{kolton_dep_zeroT_long}, showing that the depinning
geometry indeed dominates the large scales of the interface, regardless of the
magnitude of the driving force, as has been originally suggested by functional
renormalization group arguments~\cite{chauve2000}.

\begin{figure}
\includegraphics[width=\columnwidth,clip=true]{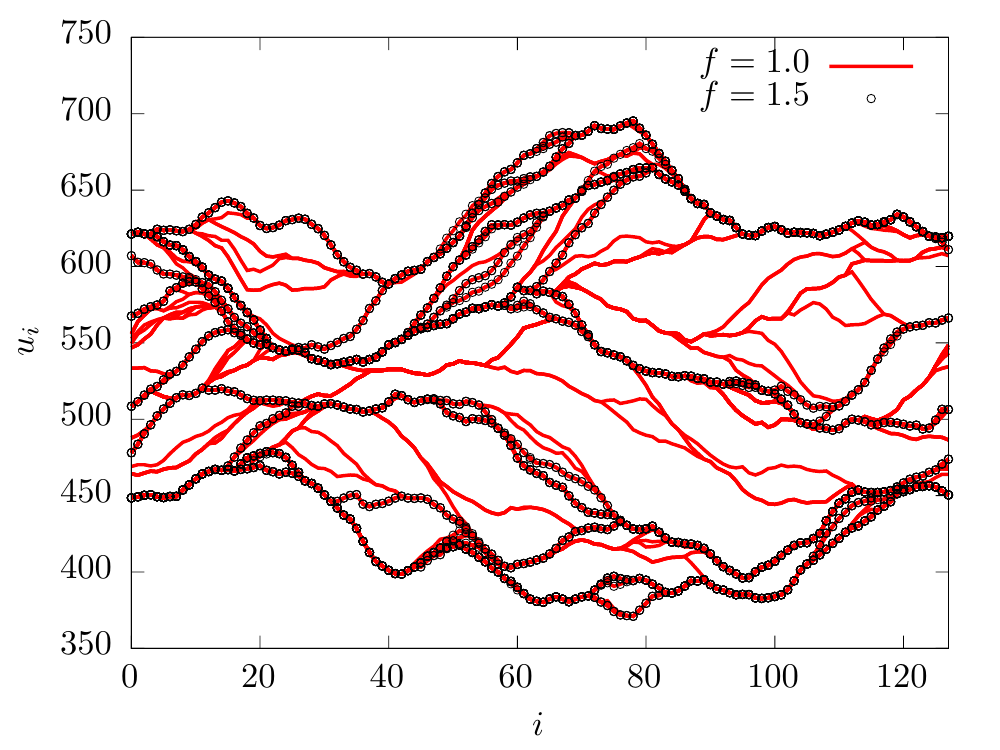}
\caption{From bottom to top: sequence of metastable
configurations visited by an elastic string of size $L=128$ driven upwards by
the subthreshold forces $f_2=1.0$ (red lines) and $f_1=1.5$ (black circles), in
the low-temperature limit for the same realization of the disorder. The initial
configuration, at the bottom, is the same for both sequences, and is an
arbitrary metastable state prepared at $f_1 < f_c \approx 1.6$. We find that, in
general, any metastable state of the $f_1$-sequence is contained in the
$f_2$-sequence, provided that $f_2<f_1$. It follows that the critical
configuration itself belongs to any $f$-sequence, provided $f<f_c$.}
\label{fig:middletoncreep}
\end{figure}

Let us consider a state with all $x_i>f$ ($f$ being the applied force). This is
a metastable state of the system and corresponds to a pinned configuration of
the interface. A vanishingly small temperature acts by eventually  destabilizing
the less stable site (the one with the lowest $x_i$), starting an avalanche that
takes the system to a new pinned configuration, with all $x_i>f$. The evolution
of the interface can thus be described as a sequence of metastable
configurations. Some general properties of the evolution of elastic interfaces
(generally known as Middleton rules~\cite{middleton_rules}) allow us to prove
the following property: given two subcritical values of the applied force $f_1$
and $f_2<f_1$, the sequence of metastable configurations that occur under $f_1$
is a subsequence of that occurring under $f_2$. A detailed demonstration of this
statement will be presented elsewhere, now we simply illustrate this fact in
Fig.~\ref{fig:middletoncreep}, where we see indeed how it takes place. In
particular, this result allows us to show that under the action of {\em any}
$f_2<f_c$, the system will eventually pass through  (meta)-stable configurations
of any $f_1>f_2$ ($f_1<f_c$). Letting $f_1\to f_c$, this indicates that the
system eventually explores critical (depinning) configurations under the action
of any applied force.

\begin{figure}
\includegraphics[width=0.9\columnwidth,clip=true]{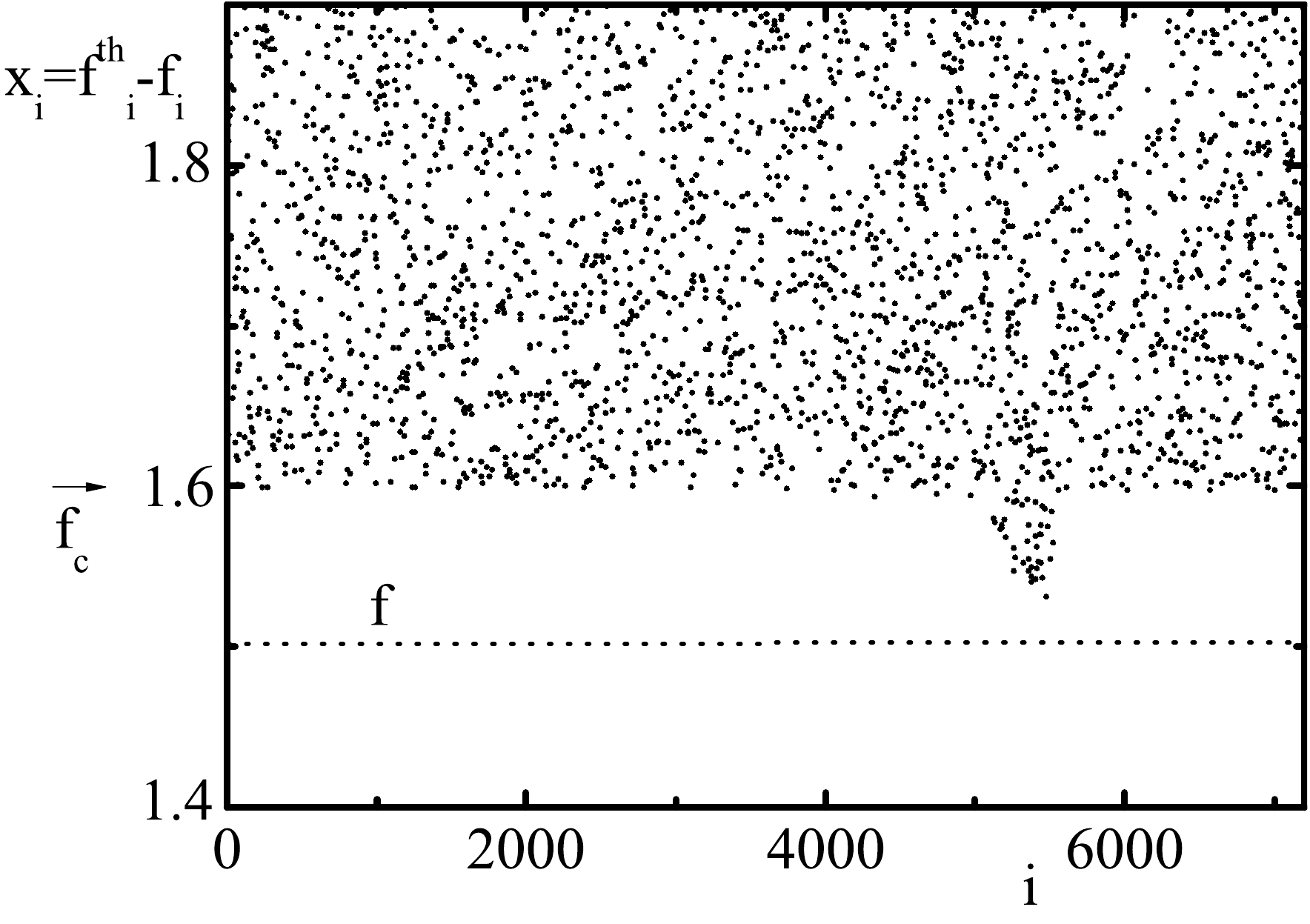}
\caption{Distribution of values of $x_i$ in the system after a long
equilibration time, using the thermal activation protocol, in the presence of an
external force $f$. Note the appearance in most of the system of a ``gap''
between $f$ and $f_c$. Sites around $i\sim 5500$ have lower values of $x_i$
because of recent avalanches in this region. It is clear that  next avalanches
will occur around this region too, i.e., highly correlated spatially with
previous avalanches.}
\label{fig:gap_thermal}
\end{figure}

Next avalanche generates values of $x_i$ above $f$, but not necessarily above
$f_c$ in the region spanned by the avalanche. The situation is shown in
Fig.~\ref{fig:gap_thermal}. We see, in fact, that in most of the system the
value of $x_i$ is above $f_c$, and only a limited region around $i\sim 5500$
(that has been affected by previous avalanches) has  abundant values of $x_i$
lower than $f_c$. It is thus clear that further avalanches will be nucleated
around the position of the previous one, leading to a strong spatial correlation
among them.

\begin{figure}
\includegraphics[width=0.9\columnwidth,clip=true]{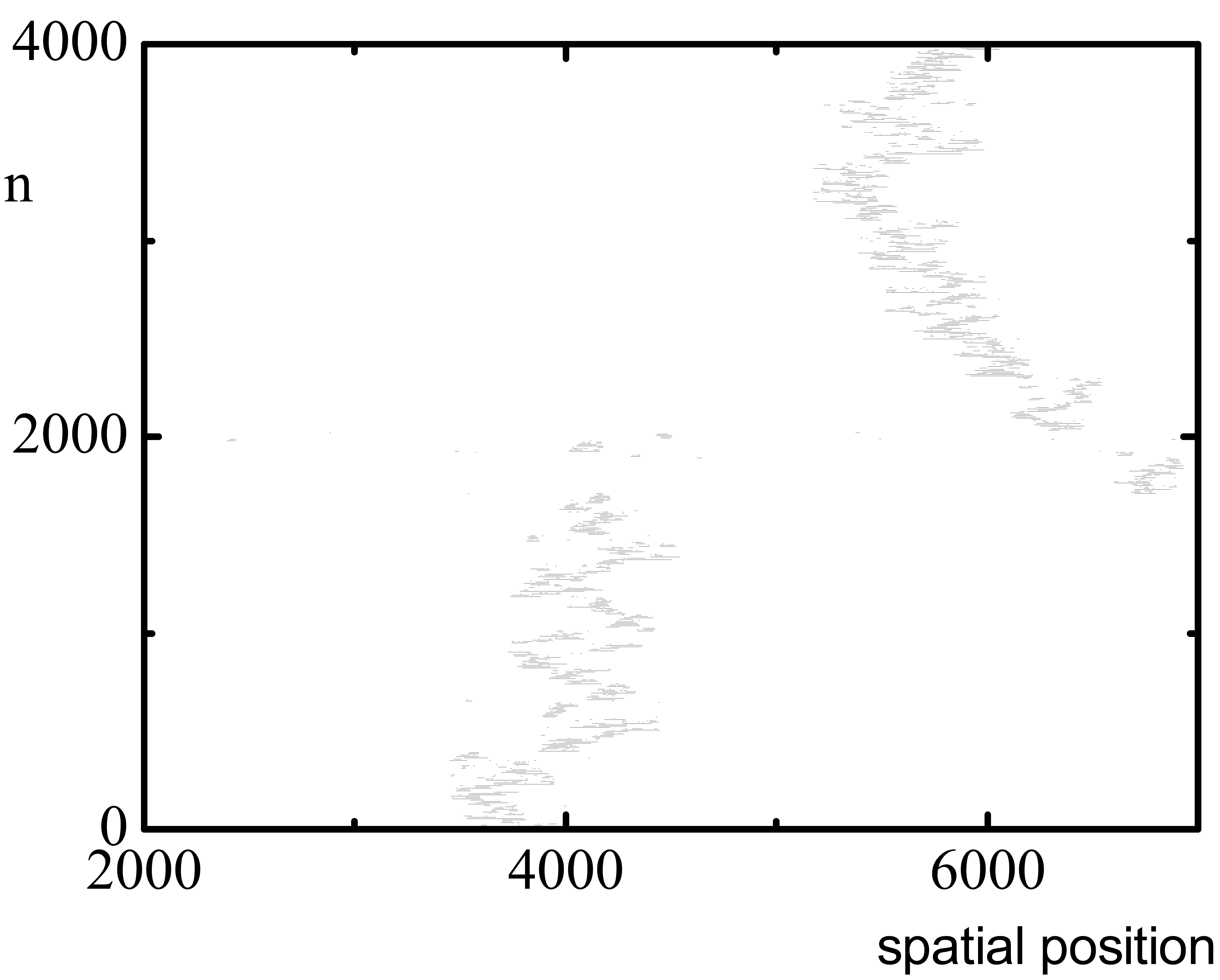}
\caption{Sequential index $n$ against the spatial extent of the avalanche. Note
the strong spatial correlation between consecutive avalanches.}
\label{fig:spatial}
\end{figure}

\begin{figure}
\includegraphics[width=0.9\columnwidth,clip=true]{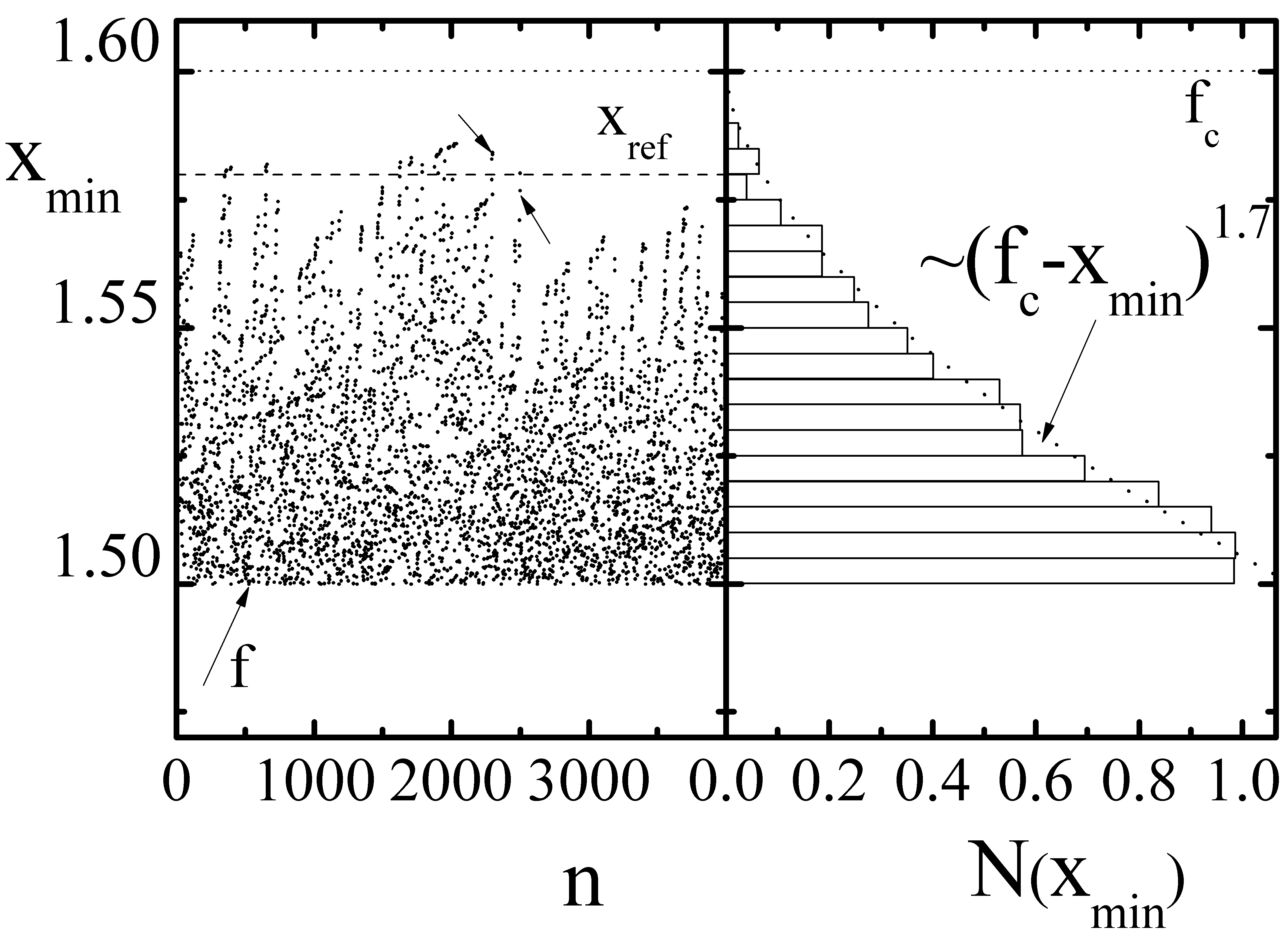}
\caption{Left: The value of $x_{\text{min}}$ for the sequence of avalanches in
the previous figure. From these values of $x_{\text{min}}$, the values of time
in the next figure are obtained as indicated in the text. Dashed line is a
reference value used to define clusters. With this threshold, the arrows point
to the first and last avalanche that form a particular cluster. Right: histogram
of the observed values of $x_{\text{min}}$. The expected exponent for the power
law is $2\nu-1$, and this coincides with the observed value.}
\label{fig:xmin}
\end{figure}

An example of the spatial correlations that appear among avalanches is shown in
Fig.~\ref{fig:spatial}. There, each avalanche is represented by a horizontal
segment covering the affected sites, the vertical coordinate being a sequential
index of the occurrence of avalanches. We clearly see the spatial superposition
between most consecutive avalanches. To appreciate also the temporal
correlation, we must calculate a realistic time to follow the process. This can
be done by first calculating the values $x_{\text{min}}$ at which each avalanche
nucleates. These values, for the sequence of 4000 avalanches in
Fig.~\ref{fig:spatial}, are shown in Fig.~\ref{fig:xmin}. Considering that the
triggering of every avalanche occurs through thermal activation over a barrier
$\epsilon\sim (x_{\text{min}}-f)^\alpha$, we can say that the waiting time
$\delta t$ for the activation of an avalanche with a given $x_{\text{min}}$ will
be $\delta t \sim \exp [(x_{\text{min}}-f)^\alpha/T]$, where the temperature
$T$ (that has been properly rescaled) is supposed to be much smaller than
typical values $(x_i-f)^\alpha$ (in our case $T\ll 0.1$).

\begin{figure}
\includegraphics[width=0.9\columnwidth,clip=true]{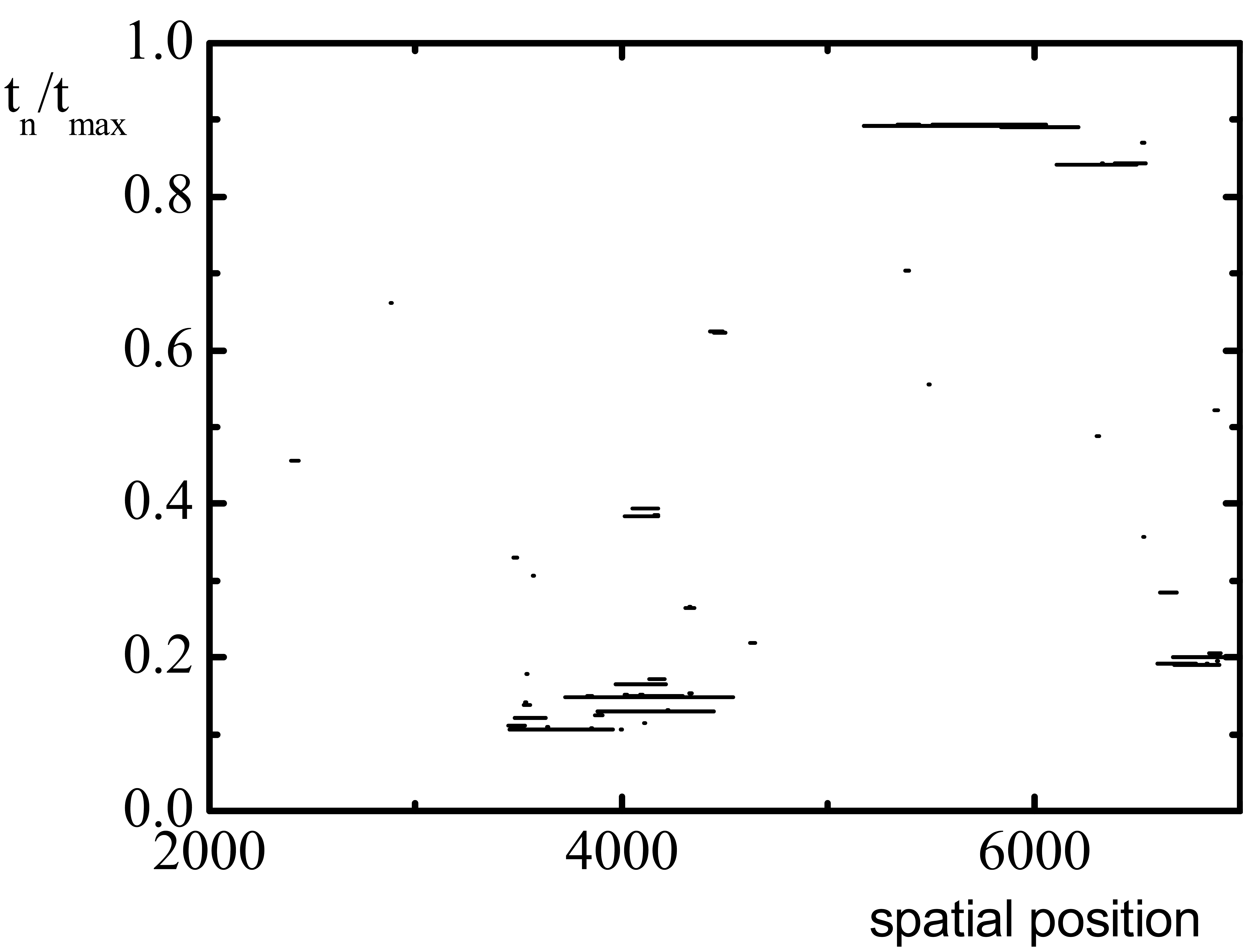}
\caption{Time of the $n$th avalanche $t_n$ (normalized to the maximum value
$t_{\text{max}}$) against spatial extent of the avalanches, obtained for
$T=0.002$, and using $\alpha=1$. The largest values of $x_{\text{min}}$ in the
previous figure are responsible of the largest temporal intervals between
consecutive avalanches in this figure.}
\label{fig:spatial_norm}
\end{figure}

Results of this calculation are presented in Fig.~\ref{fig:spatial_norm}. This
figure very naturally suggests the consideration of ``clusters'' of avalanches,
as groups of avalanches that occur close in time and space~\footnote{It must be
emphasized, however, that the concept of clusters in the present model is
qualitatively different from that in the seismic context (see Ref.~\cite{eq3}),
where they are originated in an internal relaxation mechanism, and where they
can be univocously identified in the limit of slow enough driving. In the
present case clusters are constructed using the ad hoc value of
$x_{\text{ref}}$, and cannot be defined in the absence of this reference
value.}. To define clusters, we must choose a criteria to group avalanches. In
Ref.~\cite{ferrero_correlatedcreep}, a criteria of spatial overlapping has been
used. Here we use a temporal criteria that corresponds to the one originally
used to identify avalanches in the BS model: Two consecutive avalanches belong
to the same cluster if they are separated in time less than some reference value
$t_{\text{ref}}$. From the previous discussion it is clear that this corresponds
to ask if the value of $x_{\text{min}}$ at which the second avalanche nucleates
is smaller than $x_{\text{ref}}\equiv  f+\left [T\ln(t_{\text{ref}})\right
]^{1/\alpha}$. Namely, setting a value of $x_{\text{ref}}$ (see
Fig.~\ref{fig:xmin}), clusters are formed starting at  an avalanche with
$x_{\text{min}}>x_{\text{ref}}$, and ending in the last avalanche before the
next $x_{\text{min}}>x_{\text{ref}}$. Note that it must be $x_{\text{ref}}\leq
f_c$.

Defining clusters through the introduction of the reference value
$x_{\text{ref}}$, the comparison of cluster statistics with  the avalanche
statistics at depinning is now straightforward in view of the previous
arguments. The value of $x_{\text{ref}}$ can be considered to be the value of
$f_1$ in the argument of the previous paragraphs, and it leads to the conclusion
that avalanches for any given $f<f_c$ form clusters that distribute in the same
way that avalanches at depinning as $x_{\text{ref}}\to f_c$. In particular the
size of clusters $S_{\text{cl}}$ distribute according to $P(S_{\text{cl}})\sim
S_{\text{cl}}^ {-\tau}$, with $\tau\simeq 1.11$ (in one spatial
dimension)~\cite{ferrero_correlatedcreep}.

In general terms, the issue we have just discussed concerns the relation between
the behavior of dynamical systems under parallel or extremal dynamics. This
point has been addressed also in the case of the BS model. In
Ref.~\cite{sornette} it was shown that a parallel update in the BS model
transforms it into a directed percolation problem. However, in that case, due
to the absence of an equivalent to the Middleton rules, the avalanches
generated using parallel dynamics are not equivalent to clusters of avalanches
with the extremal protocol. In fact, even the value of the critical threshold
is different in both cases: for the BS model the threshold sets at $x^*\simeq
0.6670$, whereas for the parallel update, the threshold for the corresponding DP
model is $x^*\simeq 0.5371$~\cite{sornette}. In addition,
Grassberger~\cite{grassberger} has numerically shown that critical exponents of
BS and DP do not coincide, indicating that parallel and extremal update produce
critical behavior in two different universality classes in that case.

\subsection{Velocity near the depinning threshold}
\label{sec:depinning}

The previous analysis, and in particular the identification of typical $x_i$
values in the system that display a gap between $f$ and $f_c$, allows us to
examine in more detail the asymptotic form of the velocity in the limit of very
low temperature. We can calculate the velocity $v$ of the system in this limit
as the ratio between a given advance of the interface $u_0$ (which will be taken
as a fixed parameter), and the time $t_0$ necessary  to obtain such an advance.
During the advance, different values of $x_{\text{min}}$ are encountered and
have to be activated, the value of $t_0$ being thus equal to the sum over all
$x_{\text{min}}$ values involved.
A site with a given value $x_i$ takes typically a time $t_i \sim
\exp[(x_i-f)^\alpha/T]$ to be activated. The minimum $x_i$, namely
$x_{\text{min}}$, is activated in the minimum time $t_{\text{min}} \sim
\exp[(x_{\text{min}}-f)^\alpha/T]$. However, as there are many sites with $x_i$
close to $x_{\text{min}}$, the first activation time is reduced, in the same way
that the first occurrence of one of $M$ Poisson processes with the same rate $r$
is $r/M$, instead of being simply $r$. For a fixed temperature, the sites that
have a comparable probability to become active are those with $x_i$ between
$x_{\text{min}}$ and $x_{\text{min}}+T^{1/\alpha}$. Namely, the time to be
expected up to the next activation is typically given by
$T^{-1/\alpha}\exp[(x_{\text{min}}-f)^\alpha/T]$. Now, we sum over all possible
values of $x_{\text{min}}$ to obtain $t_0$, i.e.,
\begin{equation}
t_0 \sim T^{-1/\alpha} \int _f^{f_c} dx_{\text{min}} N(x_{\text{min}})\exp
[(x_{\text{min}}-f)^\alpha/T],
\label{eq:t_0}
\end{equation}
where $N(x_{\text{min}})dx_{\text{min}}$ represents the number of times that
values between $x_{\text{min}}$ and $x_{\text{min}}+dx_{\text{min}}$ are found
along the temporal evolution, i.e., the distribution of $x_{\text{min}}$ values
observed, for instance, in Fig.~\ref{fig:xmin}. This distribution depends on the
value of $f$, but is independent of $T$ since in this limit of very low
temperature, $T$ enters only in determining the time scale of the dynamics.

As shown in Ref.~\cite{Paczuski_1996}, the distribution of $N(x_{\text{min}})$
is related to the average size of avalanches as a function of the applied force
$\overline S(f)$. In concrete, they find $N(x_{\text{min}})\sim (\partial
\overline S(f)/\partial f)^{-1}$. Taking into account that $\overline S(f)\sim
S_{\text{max}}(f)^{2-\tau}\sim (f_c-f)^{-\nu(1+\zeta)(2-\tau)}$, and using that
$\tau=2-2/(1+\zeta)$ for qEW (both for the size-distribution of deterministic
avalanches at $f=f_c$~\cite{Rosso_2009} and activated events clusters at
$f<f_c$~\cite{ferrero_correlatedcreep}), we finally obtain
\begin{equation}
N(x_{\text{min}})\sim (f_c-x_{\text{min}})^{2\nu-1}.
\label{eq:pxmin}
\end{equation}
As shown in Fig.~\ref{fig:xmin}, this result is nicely verified in the numerics.

Inserting this result into Eq.~(\ref{eq:t_0}) we obtain
\begin{equation}
t_0 \sim T^{-1/\alpha} \int _f^{f_c} dx_{\text{min}} (f_c-x_{\text{min}})^{2\nu-1}\exp
[(x_{\text{min}}-f)^\alpha/T].
\end{equation}

Calculating the leading order of the integral, we obtain the velocity of the
interface at very small $T$ and negative $\Delta=f-f_c$ as
\begin{equation}
v(\Delta,T) \sim {T^{-2\nu+1/\alpha}}{|\Delta|^{2\nu(\alpha-1)}} \exp
(-|\Delta|^\alpha/T),
\label{eq:v_lim}
\end{equation}
which is similar, for $\alpha=1$, to the one found in Ref.~\cite{vandembroucq}
using an extremal dynamics model.

\begin{figure}
\includegraphics[width=\columnwidth,clip=true]{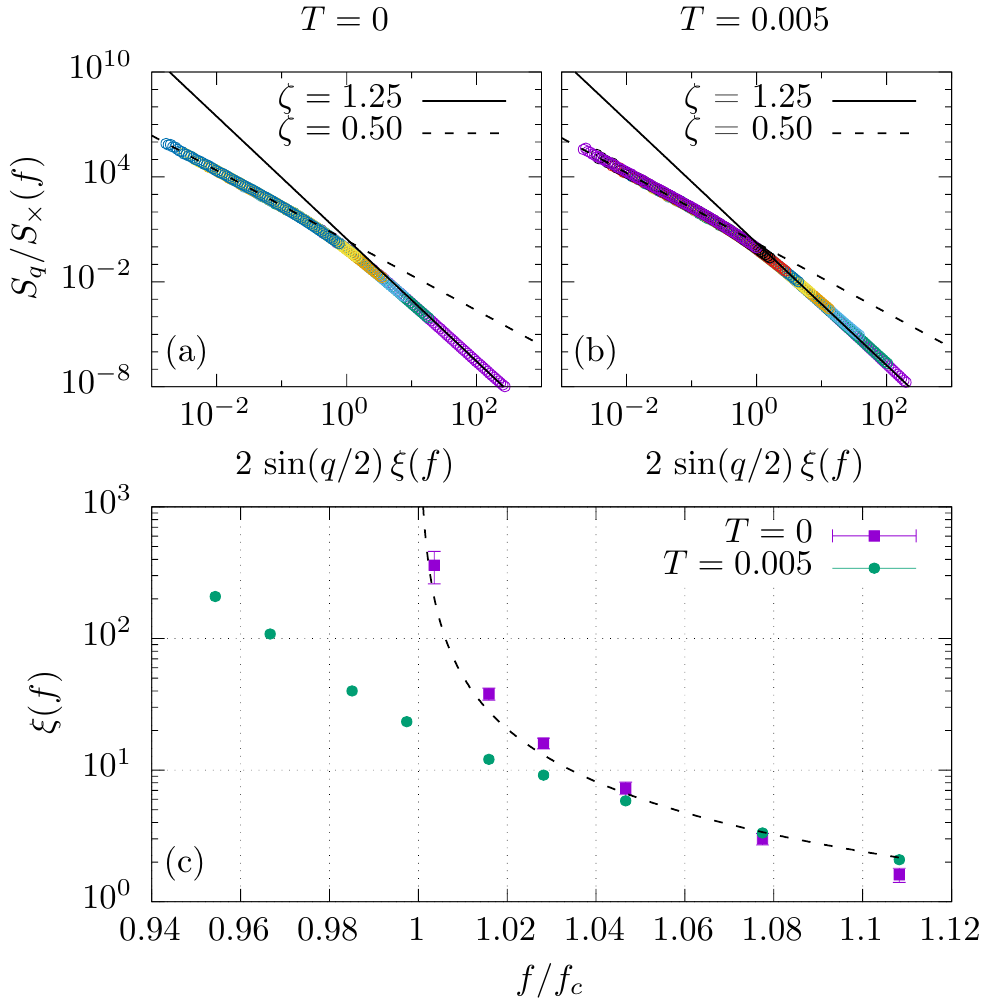}
\caption{Spatial correlations around the depinning transition.
The steady-state  disorder averaged structure factor $S_q$ at $T=0$ (a) and
$T=0.005$ (b) displays a roughness crossover, between $S_q \sim 1/q^{1+2\zeta}$
with $\zeta\approx 1.25$ and $S_q \sim 1/q^2$ at the characteristic length-scale
$\xi(f,T)$. (c) Force dependence of $\xi(f,T)$ for the two temperatures. The
dashed-line indicates the expected divergent behavior of the $T=0$ depinning
transition, $\xi(f,T=0) \sim (f-f_c)^{-\nu}$, with $\nu = 1/(2-\zeta) \approx
1.33$. At finite temperature the divergence of $\xi$ at $f_c$ disappears, and
$\xi(f,T)$ monotonically grows with decreasing $f$ in the creep regime. In
either case, the large-scale geometry is described by the critical roughness
$\zeta$.}
\label{fig:sqcreep}
\end{figure}

Summarizing the results in this section, in the creep regime in which
temperature is vanishingly small, the system evolves toward a configuration in
which typically all sites have a value of $x_i>f_c$. This means that $f$ can be
increased up to $f_c$ without triggering any avalanche in the system, and this
demonstrates that this configuration is actually a critical depinning one.
Moreover, the velocity of the interface [see Eq.~(\ref{eq:v_lim})] has a
thermally activated form, with $|\Delta|^\alpha$ being the maximum barrier
height to be overcome, and a power-of-$T$ and $\Delta$ prefactor originated in
the distribution of values of $x_{\text{min}}$ that are activated along the time
evolution.

The dominance of the maximum barrier $|\Delta|^\alpha$ (which is originated in
its exponential influence in the activation times) allows us to claim that the
critical configuration, which we showed is always visited for any applied force
$f<f_c$, becomes also the dominant state for any $f<f_c$. Notably, the dominance
of the depinning critical state goes beyond the $T\to 0+$ limit of the creep
dynamics: it dominates the large-scale geometry in the whole creep regime at any
finite temperature. To confirm this for our model we have computed the
steady-state disorder averaged structure factor $S_q \equiv \langle |u_q|^2
\rangle$ for various forces around $f_c$ both at $T=0$ and $T>0$. In
Figs.~\ref{fig:sqcreep}(a) and \ref{fig:sqcreep}(b), we see that at large
scales $S_q \sim q^{-(1+2\zeta)}$ with $\zeta\approx0.5$, reflecting the
dominance of the fast-flow geometry of the qEW model at large length scales,
only provided that the velocity is finite. Below a characteristic scale
$\xi(f,T)$ we observe a crossover toward $\zeta \approx 1.25$, the depinning
roughness exponent of the qEW class. The behavior of $\xi(f,T)$ is shown in
Fig.~\ref{fig:sqcreep}(c) at zero and finite temperature. At $T=0$, $\xi(f,T)$
tends to diverge at $f_c$, following closely the expected critical behavior of
the depinning correlation length, $\xi(f,T=0)\sim (f-f_c)^{-\nu}$ with
$\nu=1/(2-\zeta) \approx 1.33$. At finite temperature the divergence at $f=f_c$
disappears, and $\xi(f,T>0)$ grows monotonically with decreasing the force. This
crossover was also observed in molecular dynamics simulations of the smooth qEW
model, between the case $T=0$ for $f>f_c$~\cite{duemmer_2005}, and at $T>0$ and
$f \lesssim f_c$~\cite{kolton_dep_zeroT_long}. Interestingly, however, our model
displays the same geometric crossover without actually having a proper fast-flow
regime $v \sim f$ when $f \gg f_c$, because the velocity saturates in this
limit. The present results obtained in the creep regime are consistent with the
``depinning-like'' features unveiled by functional renormalization group
calculations at scales larger than the activation scale~\cite{chauve2000}, and
with the numerically observed depinning roughness at such scales, both at
intermediate~\cite{kolton_dep_zeroT_long} and vanishing driving forces in the
creep regime~\cite{ferrero_correlatedcreep}. Quite interestingly, on the other
hand, the similarity of our model at low temperatures below the depinning
threshold with Zaitsev's model~\cite{Zaitsev_1992} suggests a possible
nontrivial connection, at a \textit{coarse-grained} level, between
\textit{collective} creep and self-organized criticality.

\section{Generalized heuristic thermal rounding scaling}

We may check the analytical form of the velocity in the low-temperature regime
and for $f<f_c$ [see Eq.~(\ref{eq:v_lim})] against the numerical simulations.
Equation~(\ref{eq:v_lim}) for $\alpha=1$ is plotted as red lines on top of the
data in Fig.~\ref{fig:th0}, adjusting only a single global factor, the same for
all curves. We see that the numerical data adjust to this behavior very well,
pointing to both the accuracy of the numerical simulations and the correctness
of the analytical expression.

We have already shown that the results of numerical simulations seem to be
incompatible with the scaling proposed in Eq.~(\ref{eq:v_t_f}). Now we can
confirm this from the analytical expression we have obtained in the creep
regime: Eq.~(\ref{eq:v_lim}) (for $\alpha=1$, for instance) is not of the form
of Eq.~(\ref{eq:v_t_f}). The failure of the scaling due to the activated
behavior at very low temperature had already been noticed in
Ref.~\cite{vandembroucq}.

Yet, the results obtained up to here (both numerical and analytical) allow us to
propose heuristically a form for the thermal scaling. We restrict to the
case $\alpha=1$, which is the one for which we have obtained more reliable
numerical data. On one side the form of the plot in Fig.~\ref{fig:otrosc}(a)
suggests that a $T$-dependent horizontal shift of the curves may scale all of
them onto a universal one. On the other side, the form of Eq.~(\ref{eq:v_lim})
suggests (for $\alpha=1$) that this shift may have origin in the $T^{-2\nu+1}$
pre-exponential factor. Let us write Eq.~(\ref{eq:v_lim}) in the form
\begin{equation}
v(\Delta,T) \sim T^\beta \exp \left [\Delta /T-(2\nu-1+\beta)\ln(T/T_0)\right],
\end{equation}
with some unknown temperature scale $T_0$. Our proposal is that the previous
form, which is valid for $f<f_c$ and $T\ll f_c-f$, may be extended to a full
scaling valid for both negative and positive $f_c-f$ and $T\ll f_c$ by
generalizing the exponential function to a single scaling function $g$, in such
a way that
\begin{equation}
v(\Delta,T) \sim T^\beta g \left [\Delta/T-(2\nu-1+\beta)\ln(T/T_0)\right].
\label{eq:new_sc}
\end{equation}
For $x\to -\infty$, the $g(x)$ function behaves as $\exp(x)$. On the other hand,
for $x\to +\infty$ we expect $g(x)\to x^\beta$, in such a way that for
$\Delta=0$ (i.e., $f=f_c$) we obtain
\begin{equation}
v(0,T) \sim T^\beta g [-(2\nu-1+\beta)\ln(T/T_0)]\sim [-T\ln(T/T_0)]^\beta,
\end{equation}
i.e., a logarithmically corrected power law.

\begin{figure}
\includegraphics[width=0.9\columnwidth,clip=true]{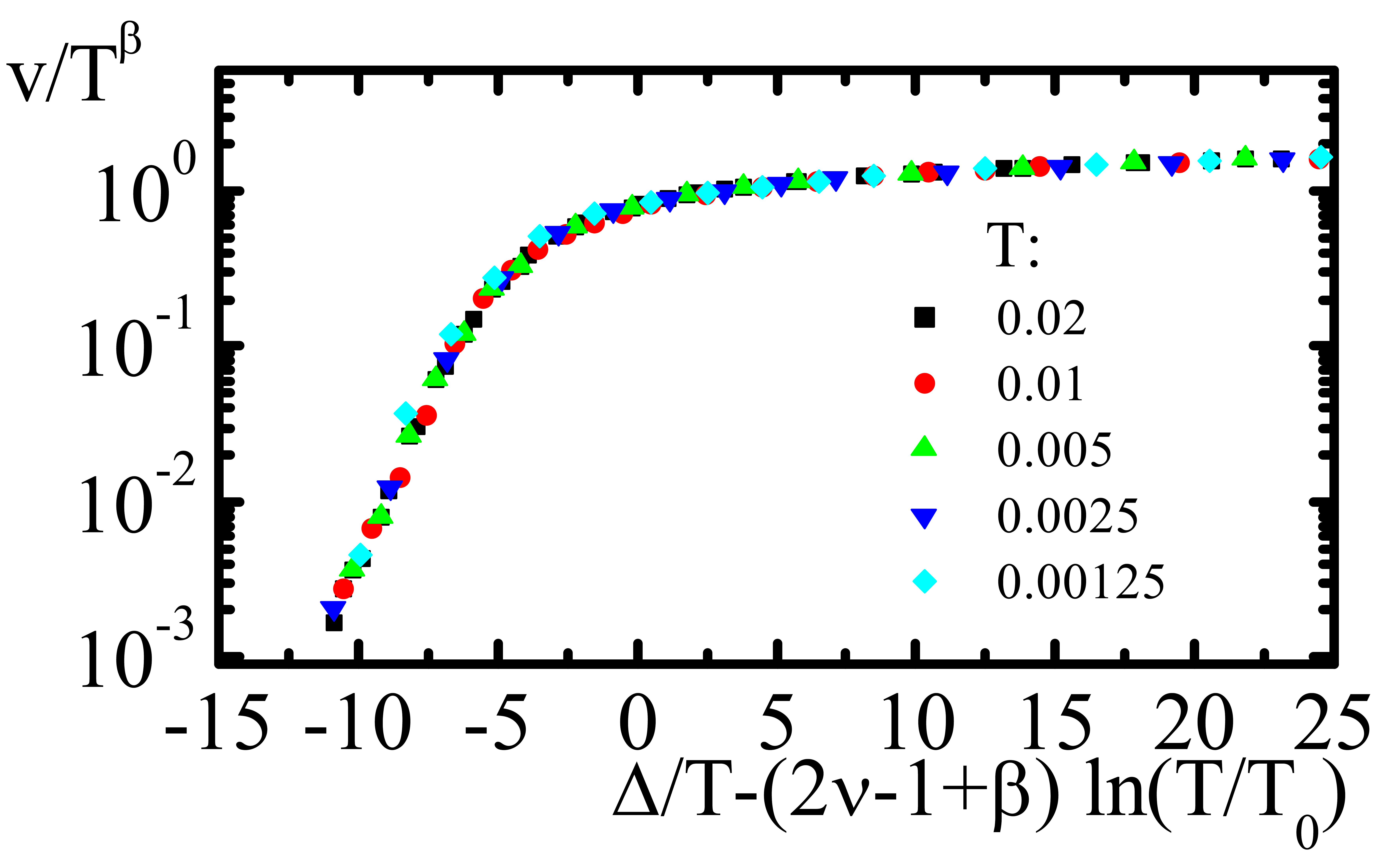}
\caption{Generalized scaling of the $v(\Delta,T)$ data. Here, the
value $T_0=0.1$ is used, but note anyway that any value of $T_0$ would be
equally effective.}
\label{fig:new_scaling}
\end{figure}

We obtain support for this conjecture from the numerics. The result of plotting
the data in Fig.~\ref{fig:th0} according to Eq.~(\ref{eq:new_sc}) is presented
in Fig.~\ref{fig:new_scaling}, where we see that the existence of a single
scaling function $g$ is well supported.

\begin{figure}
\includegraphics[width=0.9\columnwidth,clip=true]{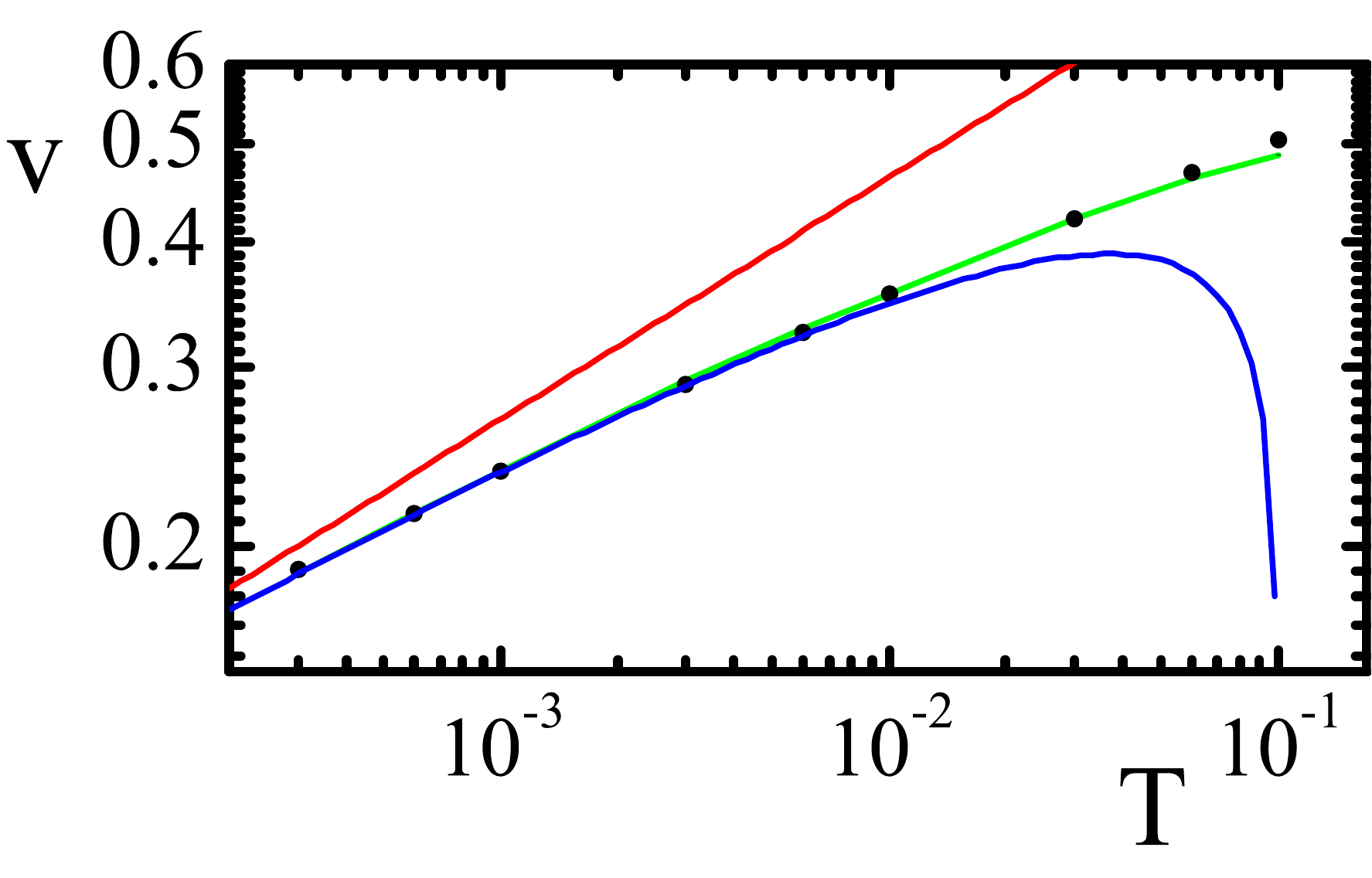}
\caption{Symbols reproduce the same data as in Fig.~\ref{fig:psitilde2}. Red
line is $\sim T^\beta$, and blue one is the expected asymptotic form $\sim
[-T\ln(T/T_0)]^\beta$, for $T_0=0.1$. Green line is the result expected using
the full form of the function $g(x)$.}
\label{fig:corr_to_scal}
\end{figure}

Additionally, in Fig.~\ref{fig:corr_to_scal} we present again the data at
$f=f_c$. The red line is the bare $T^\psi$ power that we knew already that does
not fit the results. Blue line is the asymptotic form $\sim
[-T\ln(T/T_0)]^\beta$ we expect according to the new scaling. Good agreement
with the simulated points can be achieved for the lowest values of $T$. For
larger $T$, the asymptotic form deviates from the simulated points, but we can
show that the simulated values are consistent with the full scaling. For each
value of $T$, we calculate $x\equiv -(2\nu-1+\beta)\ln(T)$, and calculate $g(x)$
from the plot in Fig.~\ref{fig:new_scaling}. Finally, we calculate $T^\beta
g(x)$ and plot the results as a green line in Fig.~\ref{fig:corr_to_scal}. We
see that the results nicely match the simulated values.

At this point, beyond the overall numerical consistency,  we have to mention
that we do not have a convincing argument explaining why the present scaling
should be appropriate. But the numerical evidence just presented strongly
supports the possibility that logarithmic corrections appear in the temperature
dependence of the data, in particular, in the form of $v(\Delta=0,T)$ [see
Eq.~(\ref{eq:new_sc})], and this can strongly affect a naive determination of
the thermal rounding exponent $\psi$.

In the theory of critical phenomena in equilibrium phase transitions, the
possibility of logarithmic corrections is well known, and a consistent theory of
their appearance is available~\cite{kenna}. One of the key points that such
analysis reveal, is that logarithmic corrections are expected only in
particular circumstances. One is the case in which the dimensionality of the
system corresponds to the critical dimension of the problem separating
mean-field and nontrivial scaling behavior. It is thus natural to ask if the
behavior we obtain for the case of thermal rounding is also tuned to
dimensionality. Although we did not address this problem yet, our expectation is
that logarithmic correction for thermal activation should also be present in any
dimension, since these corrections appear in the present case as a consequence
of the dependence of the activation time on temperature for single sites, i.e.,
they are not related to some crucial feature associated to dimensionality.

\section{Conclusions}

Using a model with random traps, we have shown that the slow velocity regime
near the depinning threshold of an elastic interface in a random medium is very
different for uniform stochastic activation than for Arrhenius thermal
activation. In the former case, the velocity accurately follows a standard
scaling law involving force and noise intensity, with the analog of the thermal
rounding exponent satisfying a slightly modified ``hyperscaling'' relation. It
would be interesting to investigate if there is some formal connection between
the relation we derive here and standard hyperscaling relations that are valid
only at equilibrium. For the Arrhenius activation, we find instead that standard
scaling fails  for \textit{any} value of the thermal rounding exponent and
propose a modified form to satisfactorily describe the data. We argue that the
anomalous scaling of the velocity is related to the strong correlation existing
between activated hops, which, alternated with deterministic depinning-like
avalanches, occur below the depinning threshold. We rationalize this
spatiotemporal patterns---interestingly very similar to the ones reported for
collectively activated events in creep simulations at low driving
forces~\cite{ferrero_correlatedcreep}---by making an analogy of the present
model in the near-threshold creep regime with some well-known models with
extremal dynamics, particularly the Bak-Sneppen model.

We hope the present results will motivate further research on the creep and
thermal rounding regimes of elastic manifolds in random media. In this respect,
we note that the thermal rounding regime was recently observed experimentally in
thin-film ferromagnets~\cite{jeudy_PRL_enviado}. In that case, restricting to
data \textit{above} the estimated critical depinning field, a scaling function
and an effective thermal rounding exponent $\psi \approx 0.15$ were found. It
would be thus interesting to see if the kind of scaling functions we propose
provide a better collapse, \textit{both} below and above the depinning
threshold, as we observe in our simulations. Furthermore, it would be very
interesting to experimentally measure the spatiotemporal patterns we observe
for the activated events, which are similar to the ones reported
in Ref.~\cite{ferrero_correlatedcreep} by using a different model.

\begin{acknowledgments}

We thank S. Bustingorry and E. Ferrero for helpful discussions. We acknowledge
partial support from Grants No. PICT 2012-3032 (ANPCyT, Argentina) and No. PIP
2014-0681 (CONICET, Argentina). This work used Mendieta Cluster from CCAD-UNC,
which is part of SNCAD-MinCyT, Argentina.

\end{acknowledgments}

\appendix
\section{Thermal rounding scaling for one particle}
\label{sec:single_par}

In this appendix, we analyze the scaling of the velocity force characteristics
$v(f,T)$ for a pinned Brownian particle in a ring. We consider a simple one
parameter family of pinning potentials yielding different critical exponents.
For this family, we first derive $\beta$, $\psi$ and the barrier exponent
$\alpha$ such that above the threshold $v(f,T=0)\sim (f-f_c)^\beta$, at the
threshold $v(f=f_c,T) \sim T^{\psi}$, and below the threshold $v(f \lesssim
f_c,T) \sim \exp[-(f_c-f)^\alpha/T]$, provided $T \ll f_c a$ (with the
periodicity fixed to $a=1$ from now on). With these exponents, we test the
scaling prediction of Eq.~(\ref{eq:v_t_f}).

Let us consider the overdamped Langevin dynamics,
\begin{equation}
\dot{u} = f + F(u)+\eta(t),
\label{eq:langevin}
\end{equation}
where $\eta(t)$ is a Langevin noise at temperature $T$, characterized by
$\langle \eta(t) \rangle=0$,  $\langle \eta(t)\eta(t') \rangle=2T\delta(t-t')$.
We will assume a periodic pinning force, $F(u)=F(u+n)$ with $n$ any integer,
with a well-defined steady-state depinning transition at $f = f_c \equiv
-\max_{u} \{ F(u) \} = -F(u_c)$, with $u_c$ the critical position. To fix ideas,
we will assume that at $T=0$ and $f<f_c$ there are only two fixed points, one
stable and the other unstable, that annihilate exactly at $f=f_c$, at the
critical depinning position $u_c$. For $f \gtrsim f_c$ there are no fixed points
and the particle will spend most of its time around the bottleneck created at
$u_c$. This bottleneck will dominate the critical behavior of the finite mean
velocity. A general normal form for the saddle-node bifurcation described above
is
\begin{equation}
\dot{u} \approx (f-f_c)+|u-u_c|^{\gamma},
\label{eq:normalform}
\end{equation}
actually representing a family parameterized by the exponent $\gamma$. Quite
generically, smooth forces can be developed to second order near their minimum
at $u_{\text{min}}$ as $F(u)\sim F(u_{\text{min}})
+F''(u_{\text{min}})(u-u_{\text{min}})^2/2 + {\cal O}[(u-u_{\text{min}})^3]$, so
they correspond to the ``standard'' case $\gamma=2$. The case $\gamma \neq 2$
hence represents, for a particle, an ``anomalous marginality'' case which will
be useful for comparison with the elastic string case. We will focus on the
cases $\gamma \geq 1$ where it is easy to show that the mean velocity and the
barriers to motion display critical behavior: the mean velocity behaves as $v =
[\int_0^1 du'/(f+F[u'])]^{-1} \sim (f-f_c)^\beta$ just above $f_c$, while just
below $f_c$ the the particle must overcome an energy barrier $\Delta E \equiv
-\int_{u_{\circ}}^{u_{\bullet}} [f+F(u')] du' \sim (f_c-f)^\alpha$ to
move forward, where $u_{\circ} \approx u_c - (f_c-f)^{1/\gamma}$ is the stable
fixed point of the dynamics in the unit interval, and $u_{\bullet} \approx u_c +
(f_c-f)^{1/\gamma}$ is the unstable fixed point. We will be particularly
interested in the mean velocity at $f=f_c$ where $v \sim T^{\psi}$ is expected.
Simple estimations using the normal form of Eq.~(\ref{eq:normalform}) near $f_c$
show that
\begin{eqnarray}
\beta&=&1-1/\gamma, \nonumber \\
\alpha&=&1+1/\gamma, \nonumber \\
\psi&=&\beta/\alpha =(\gamma-1)/(\gamma+1).
\label{eq:oneparticle}
\end{eqnarray}
If $\Delta E \gg T$ the velocity is controlled by Arrhenius activation $v \sim
\exp[-\Delta E/T]=\exp[-2(f_c-f)^\alpha/T]$. Interestingly, the scaling $v \sim
T^{\psi}$ with $\psi=\beta/\alpha$ coincides with the prediction for a charge
density wave system~\cite{middleton} assuming that just below the depinning
threshold the elastic manifold depins by thermally exciting localized modes,
which can then trigger large deterministic avalanches. A crucial assumption in
this picture is that the energy barrier to activate such modes scales as
$(f_c-f)^\alpha \sim T$ just below the depinning threshold. Hence, the time
scale $\sim (f_c-f)^{-\nu z}$ the avalanche evolves is much larger than the
activation time and then the velocity is controlled by the avalanche evolution,
yielding $v \sim (f_c-f)^\beta \approx T^{\beta/\alpha}$. To test this
conjecture, the value of $\alpha$ should be first determined by the microscopic
potential renormalized at such localization scale. Reasonably, the single
particle $\alpha=3/2$ has been usually conjectured, leading to the prediction
$\psi=2\beta/3$, which mixes the ``mesoscopic'' nonuniversal exponent $\alpha$
with the universal exponent $\beta$ dominated by the large-scale behavior.

\begin{figure}[!htb]
\includegraphics[width=\columnwidth,clip=true]{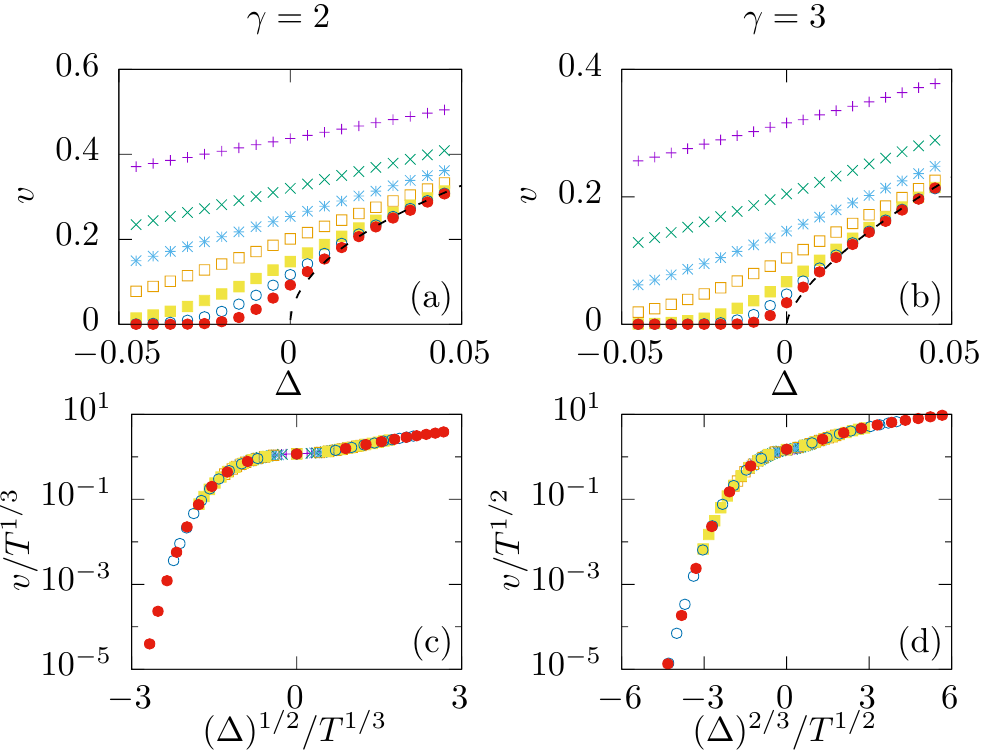}
\caption{Mean velocity as a function of the driving force and
temperature (temperature increases from the bottom to the top curves) for a
particle in a periodic potential described by Eq.~(\ref{eq:onepartpot}) with
$\gamma=2$ (a) and $\gamma=3$ (b). Both family of curves can be collapsed by
rescaling using Eq.~(\ref{eq:v_t_f}) with the appropriate critical exponents of
Eq.~(\ref{eq:oneparticle}): $\gamma=2$ (c) and $\gamma=3$ (d).}
\label{fig:onepart}
\end{figure}

To test the scaling of the velocity as a function of the drive and
temperature, we now go beyond the normal forms of Eq.~(\ref{eq:normalform}) and
propose a concrete form for $F(u)$ containing them. A simple well-known form is
$F(u) \sim -\cos(2 \pi u)$, modeling the overdamped dynamics of the
superconducting phase difference in a nonextended Josephson
junction~\cite{ambegaokar1969}. For this simple case, corresponding to a
standard $\gamma=2$ marginality, there exists an exact analytical closed
expression for the mean velocity as a function of the force and temperature, and
the values $\beta=1/2$, $\alpha=3/2$, and $\psi=1/3$ are well
known~\cite{bishop1978}. The value $\psi=1/3$ is in particular well known from
the study of thermal effects in ``spinodal decomposition''
models~\cite{Caroli1980}. The result of Eq.~(\ref{eq:oneparticle}) for $\psi$
thus generalizes this result for ``anomalous marginality,'' $\gamma \neq 2$.
Here we will use an extension of the overdamped Josephson-junction problem:
\begin{equation}
F(u)=\frac{[1-\cos (2 \pi  u)]^{\gamma/2}}{\frac{2^{\gamma/2} \Gamma
\left({\gamma/2+\frac{1}{2}}\right)}{\sqrt{\pi } \Gamma (\gamma/2+1)}}-1,
\label{eq:onepartpot}
\end{equation}
for which we have set the constant factors to have $f_c=1$, $u_c=0$ and
$\int_0^1 du\; F(u)=0$. It is clear that this model reduces the well-known
Josephson junction problem with $F(u)=-\cos(2\pi u)$ for $\gamma=2$, and the
normal form is described by the normal form family proposed in
Eq.~(\ref{eq:normalform}).

The full mean velocity characteristics $v(f,T)$ for the general
Eq.~(\ref{eq:langevin}) can be obtained in an analytical form involving two
integrals over the period of the potential~\cite{ledoussal_1995}. We can either
evaluate these integrals numerically or directly solve the Langevin equation
averaging the velocity for many noise realizations. Here we have exploited the
embarrassingly parallel nature of the latter approach, by concurrently
simulating millions of Brownian trajectories in general purpose graphical
processing units. In Fig.~\ref{fig:onepart} we plot the curves $v(f,T)$ thus
obtained and show that a perfect collapse is produced for two different values
of $\gamma$ and for driving forces both above and \textit{below} the threshold
$f_c=1$, using the $\gamma$-dependent exponents from Eq.~(\ref{eq:oneparticle}),
and the scaling form of Eq.~(\ref{eq:v_t_f}). These results are in sharp
contrast with the results we report for the elastic string, where
Eq.~(\ref{eq:v_t_f}) fails, specially for $f<f_c$. This suggests that the
anomalous thermal rounding scaling of elastic manifolds may be related to the
correlations and distributions of local thresholds induced by particle-particle
interactions in an elastic string in a 2D random landscape.

\bibliography{references.bib}
\end{document}